\documentclass{emulateapj}
\begin{document}

\def\bullet{\object{1E0657$-$56}}
\def\arcsecf{\!\!^{\prime\prime}}
\def\arcminf{\!\!^{\prime}}
\def\diff{\mathrm{d}}
\def\ngx{N_{\mathrm{x}}}
\def\ngy{N_{\mathrm{y}}}
\def\eck#1{\left\lbrack #1 \right\rbrack}
\def\eckk#1{\bigl[ #1 \bigr]}
\def\rund#1{\left( #1 \right)}
\def\abs#1{\left\vert #1 \right\vert}
\def\wave#1{\left\lbrace #1 \right\rbrace}
\def\ave#1{\left\langle #1 \right\rangle}
\def\kms{{\rm \:km\:s}^{-1}}
\def\dds{D_{\mathrm{ds}}}
\def\dd{D_{\mathrm{d}}}
\def\ds{D_{\mathrm{s}}}

\def\rxj{\object{RX~J1347.5$-$1145}}

\title{Dark Matter and Baryons in the Most X-ray Luminous and Merging Galaxy Cluster RX~J1347.5$-$1145 \altaffilmark{*}} \altaffiltext{*}{Based on
observations made with the NASA/ESA Hubble Space Telescope, obtained
at the Space Telescope Science Institute, which is operated by the
Association of Universities for Research in Astronomy, Inc., under
NASA contract NAS 5-26555. These observations are associated with
program \# 10492.  This work is also based on observations collected
at the European Southern Observatory, Chile (ESO program
078.A-0746(A)).}
\shorttitle{}
\author{Maru\v{s}a Brada\v{c}\altaffilmark{1,2,x},
Tim Schrabback\altaffilmark{3},
Thomas Erben\altaffilmark{3},
Michael McCourt\altaffilmark{1},
Evan Million\altaffilmark{1},
Adam Mantz\altaffilmark{1},
Steve Allen\altaffilmark{1},
Roger Blandford\altaffilmark{1},
Aleksi Halkola\altaffilmark{3},
Hendrik Hildebrandt\altaffilmark{3},
Marco Lombardi\altaffilmark{4,5},
Phil Marshall\altaffilmark{2},
Peter Schneider\altaffilmark{3},
Tommaso Treu\altaffilmark{2,y},
Jean-Paul Kneib\altaffilmark{6}
}
\shortauthors{Brada\v{c} et al.}
\altaffiltext{1}{Kavli Institute for Particle Astrophysics and Cosmology,
2575 Sand Hill Rd. MS29, Menlo Park, CA 94025, USA}
\altaffiltext{2}{Department of Physics, University of California, Santa Barbara, CA 93106, USA}
\altaffiltext{3}{Argelander-Institut f\"{u}r Astronomie, Auf dem H\"{u}gel 71,
D-53121 Bonn, Germany}
\altaffiltext{4}{European Southern Observatory, Karl-Schwarzschild-Str. 2, D-85748 Garching, Germany}
\altaffiltext{5}{Universit\`{a} degli Studi di Milano,  v. Celoria 16, I-20133 Milano, Italy}
\altaffiltext{6}{OAMP, Laboratoire d'Astrophysique de Marseille UMR,
6110 Traverse du Siphon, 13012 Marseille, France}
\altaffiltext{x}{Hubble Fellow}
\altaffiltext{y}{Sloan Fellow, Packard Fellow}
\email{marusa@physics.ucsb.edu}


\begin{abstract}
The galaxy cluster {\rxj} is one of the most X-ray luminous and most
massive clusters known. Its extreme mass makes it a prime target for
studying issues addressing cluster formation and cosmology. Despite
the naive expectation that mass estimation for this cluster should be
straightforward (high mass and favorable redshift make it an efficient
lens, and in addition it is bright in X-rays and appears to be in a
fairly relaxed state), some studies have reported very discrepant mass
estimates from X-ray, dynamical and gravitational lensing.  In this
paper we present new high-resolution HST/ACS and Chandra X-ray
data. The high resolution and sensitivity of ACS enabled us to detect
and quantify several new multiply imaged sources, we now use a total
of eight for the strong lensing analysis. Combining this information
with shape measurements of weak lensing sources in the central regions
of the cluster, we derive a high-resolution, absolutely-calibrated
mass map. This map provides the best available quantification of the
total mass of the central part of the cluster to date. We compare the
reconstructed mass with that inferred from the new Chandra X-ray data,
and conclude that both mass estimates agree extremely well in the
observed region, namely within 400$h_{70}^{-1}$kpc of the cluster
center. In addition we study the major baryonic components (gas and
stars) and hence derive the dark matter distribution in the center of
the cluster. We find that the dark matter and baryons are both
centered on the BCG within the uncertainties (alignment is better than
$<10\mbox{ kpc}$). We measure the corresponding 1-D profiles and find
that dark matter distribution is consistent with both NFW and cored
profiles, indicating that a more extended radial analysis is needed to
pinpoint the concentration parameter, and hence the inner slope of the
dark matter profile.

\end{abstract}


\section{Introduction}
\label{sec:intro}
Galaxy clusters have been the focus of very intense research over the
past decade. They are extremely valuable for studying the empirical
properties of dark matter, as well as for exploring the growth of
structure at the high mass tail of the mass function. The mass
distribution of galaxy clusters is particularly important for
cosmological studies because it provides a critical test of the Cold
Dark Matter (CDM) paradigm and constrain dark energy models, if their
mass and its distribution is reliably determined.  However, in some
cases the cluster mass estimates from the weak lensing, strong lensing
and X-ray measurements disagree: the origin of these discrepancies is
likely to be a combination of erroneous assumption of the cluster
potential, projection effects and complicated gas physics not taken
into account. In addition, systematic effects like cluster member
contamination and unknown redshift distribution of sources, strong
lensing image identification, temperature calibration, etc. enter the
error budget for X-ray and lensing analyses.

It was first proposed by \citet{navarro97} that the dark matter halos
on a variety of scales should follow a universal profile within the
currently accepted $\Lambda$CDM paradigm. The 3-D density distribution
of dark matter should follow $\rho_{\rm DM} \propto r^{-1}$ within a
scale radius $r_{\rm s}$ and falls of steeper at radii beyond that
($\rho_{\rm DM} \propto r^{-3}$). However recent simulations have
shown that the profile might be modified and the structural parameters
vary with redshift and mass \citep{neto07,navarro04}. In addition,
likely interaction between dark matter and baryons further complicate
the picture \citep{gnedin04, nagai05}. It is therefore crucial to
compare simulations and observations keeping the baryonic component in
mind. This will allow us to measure the slopes of dark matter
profiles, which is a critical test for our currently accepting
cosmology as well as understanding the complicated baryonic physics in
galaxy clusters. Several works have previously studied mass
distribution in number of clusters using combined strong and weak
lensing reconstruction \citep[see e.g.][]{natarajan96,kneib03,smith04,
diego05, cacciato05, jee07, limousin07} and combined strong lensing and
stellar kinematics data of the dominating central galaxy
\citep{sand07}. These approaches offer valuable extra constraints for
determining the mass distributions.

In this paper we
study the most luminous X-ray clusters known to date. The cluster
{\rxj} \citep{schindler95} at a redshift $z=0.451$ has been a subject of
intense past research. Rich datasets including X-ray
\citep{schindler95, schindler97,allen02,ettori04,gitti04, gitti07} and optical
\citep{fischer97,sahu98,cohen02,ravindranath02} as well as
observations of Sunyaev-Zel'dovich (SZ) effect
\citep{pointecouteau01,komatsu01,kitayama04} have been obtained, yet the mass
determinations based on X-ray properties, SZ effect (SZE), velocity
dispersion measurement, strong and weak lensing have many times
yielded discrepant results (see \citealp{cohen02} for a summary). In
particular the discrepancy between the dynamical mass estimate
\citep{cohen02}, early X-ray mass measurements \citep{schindler97} and
lensing results \citep{fischer97} yielded a factor of $\sim 3$
discrepancy in their mass estimates.

\citet{cohen02} suggested the cluster is likely undergoing a
major merger which would reconcile the low velocity dispersion (and
hence virial mass) measurement and its high mass as predicted by
gravitational lensing and X-ray data. If the velocity dispersion
measures predominately one component, and the X-ray gas has virialized
to the post-merger cluster mass, this can possibly explain the
data. Further evidence for a merger scenario comes also from the
optical morphology: the cluster contains two cD galaxies (see
e.g. Fig.~\ref{fig:arcs} - throughout this paper we will refer to the
brighter, western one as the BCG, and the other as the second
cD). Finally, there is a region of shocked gas in the south-east
quadrant discovered and discussed by \citet{komatsu01} (with SZ effect) and
\citet{allen02} based on X-ray observations with Chandra. We confirm the latter using the newest Chandra X-ray data.

In order to resolve the puzzle of mass estimates in \citet{bradac04b}
we studied this cluster using ground-based optical data, and for the
first time determined its mass distribution using a combined strong (multiply-imaged systems) and weak (statistical measurement using
shape information of an ensemble of background galaxies) gravitational
lensing analysis. The main drawback of that reconstruction was the
lack of clearly identified multiply imaged systems and their
redshifts. In the current paper we obtained new multi-color HST/ACS
data. The high resolution and sensitivity of ACS allowed us to
unambiguously identify many new multiply imaged systems. In this
analysis we are able to use 9 multiply imaged systems, in contrast to
the single one of
\citet{bradac04b}. Furthermore the far greater density of sources that can
be used for weak lensing (a factor of 5 improvement) has allowed us to
obtain an absolutely-calibrated mass map at high resolution. We also
present new X-ray data and compare estimates from both analyses. In
addition we describe a new automated search routine for multiply
imaged sources we are currently developing; its main advantage is that
it combines colors and information on lensing geometry of the whole
system to obtain matching pairs in a semi-automated way.

Our final mass reconstruction is again performed using a
pixelated model which combines the information from strong and weak
gravitational lensing. If spectroscopic redshifts of the multiply
imaged systems were available, this would fully eliminate the need of
using models that assume a particular shape of the gravitational
potential. However, since only one system has a spectroscopically
confirmed redshift, we use the predictive power of gravitational
lensing to predict the redshifts for the remaining systems. As
explained further in \S\ref{sec:slensing}, we therefore model strong
lensing data using parametrised models at the start to obtain the
redshifts. For the final reconstruction we however do not assume a
special shape of the gravitational potential, instead exploring a much
larger range of cluster mass distributions and profiles (provided the
estimated redshifts are reasonably reconstructed). Furthermore, we use
cluster to study the interplay between baryons
and dark matter, therefore we study separately its stellar, gas and
dark matter component.  This allows us to measure the cluster's dark
matter profile, one of the strong predictions of $\Lambda$CDM
cosmology.

This paper is structured as follows. In section~\ref{sec:data} we
describe the optical data used in this analysis. The basic image
processing and the extraction of the strong gravitational lensing data
is described in section~\ref{sec:slensing} and in
section~\ref{sec:wlensing} we describe the weak gravitational lensing
data. We present the X-ray data in section~\ref{sec:xraydata}.  We
infer the mass distribution of the cluster {\rxj} from lensing data, and
compare it with X-ray data, in section~\ref{sec:resultsw}. The
individual contributions of stars, gas, and dark matter to the total
mass of the cluster are studied in section~\ref{sec:dmbar}. The
conclusions are presented in section~\ref{sec:conclusions}. To
evaluate the angular diameter distances throughout the paper we assume
the $\Lambda$CDM cosmology with $\Omega_{\rm m} = 0.3$,
$\Omega_{\Lambda} = 0.7$, and Hubble constant $H_0 = 70 {\rm \ km \:
s^{-1}\:\mbox{Mpc}^{-1}}$.

\section{Optical observations and initial data reduction process}
 \label{sec:data}

ACS/WFC imaging of the cluster {\rxj} was carried out in Cycle 14
(proposal 10492, PI Erben) on 2006 March 9-11.  The cluster was
observed in a single (dithered) pointing and three different filters
F475W, F814W, and F850LP for 5280s (two orbits) each. The demands
placed by the lensing analysis require special care when reducing the
images. We base the reduction on the bias and flat-field corrected
flt-images (provided by the standard ACS pipeline).  For each exposure
we compute a noise model including all noise sources {\it except}
object photon noise, and update the bad pixel mask as detailed in
Marshall et al. (in preparation). We subtract the sky background
seperately in the four image quadrants to correct for the sometimes
present residual bias level.  Satellite trails are masked manually, in
order to exclude them from the coaddition.  We use the {\tt
Multidrizzle} \citep{multidrizzle} routine to align the images, correct for
geometric camera distortion, mask cosmic rays, and for coaddition.  To
register the images with the astrometric accuracy needed for the
lensing analysis, we determine the offsets among the images by
matching windowed {\tt SExtractor} \citep{sextractor} positions of
high $S/N$ objects in the individual, distortion-corrected exposures.
We calculate residual shifts and rotations using the IRAF routine {\tt
geomap}, which are then fed back into {\tt Multidrizzle}.  We use
``square'' as the final drizzling kernel, where we slightly shrink the
input pixels ($\mathrm{pixfrac}=0.9$) and set the output pixel scale
to 0.03 arcsec.  This is smaller than the original pixel scale of the
ACS/WFC detector in order to reduce the impact of resampling on the shape
measurements.  In the final coaddition, we use our noise model for
inverse variance weighting.  For further details on the data reduction
see Marshall et al. (in preparation).

\section{Strong Lensing Data Analysis and Fitting Parametrised Models}
 \label{sec:slensing}

With the advent of the high resolution HST optical images the quality
and complexity of the strong lensing data increased dramatically. Many
clusters have been observed with HST showing numerous multiply imaged
systems (the most notable example being A1689 --
\citealp{broadhurst05,halkola06}). By far the most time-consuming part of the
strong analysis is, however, to match the multiply imaged system -
i.e. identify all the multiple images that were produced for a single
source.

First steps in trying to optimise this process were done by
\citet{kneib93} and \citet{sharon05}. Their methods still involve 
significant human involvement; we are aiding the process of searching
for multiple image systems with automated color and geometry
matching. Our method is currently under development and once finished
will allow for simultaneous search of the best matching objects in the
color space as well as including the geometry of the system
(i.e. positions and orientations of the lensed images). This is a very
important step in the view of large samples of galaxy clusters that
will be discovered and targeted in the future optical surveys (such as
SNAP, LSST, DUNE and PanSTARRS).

We start by considering a simple initial guess for a cluster potential
together with a hypothesis of at least one multiple image system. We
use the image positions to determine the mass distribution of the
cluster using a simple parametric model. The image positions (at the
end of this process we use all images listed in Table~\ref{tab:arcs}
and shown in Fig.~\ref{fig:arcs}) are used in {\tt LENSTOOL}, ({\tt
http://www.oamp.fr/cosmology/lenstool/}, \citealp{kneib93}). The code
determines the best fit parameters of a parametrised model.  In
addition, its Bayes optimization and Monte Carlo Markov Chain (MCMC)
sampling routines \citep{jullo07} allow us to obtain a set of all
models that satisfactory fit the data (for few model parameters, such
as the mass and ellipticity of the cluster). We characterize {\rxj} by
two PIEMD profiles (see e.g. \citealp{limousin05,eliasdottir07}
for the basic properties of the model), centered on the two brightest
cluster members. In addition we also include the 20 brightest cluster
members in the I-band to the mass model. They are each modelled as
PIEMD spheres with a line-of-sight velocity dispersion $\sigma^{\rm
g}_{\rm PIEMD}$ and cut radius $r^{\rm g}_{\rm cut,PIEMD}$
proportional to the luminosity $L$ of each member ($\sigma^{\rm g}_{\rm PIEMD}
\propto L^{1/4}$ and $r^{\rm g}_{\rm cut,PIEMD} \propto L^{1/2}$). The best-fit scaling was found to be $\sigma^{\rm g}_{\rm PIEMD} = (260 \pm 50) 
\mbox{ km\,s}^{-1}$ and $r^{\rm g}_{\rm cut, PIEMD} = (5\pm 1) \mbox{kpc}$ for a galaxy with $m_{\rm I} = 20.5$. In total this adds to 50 constraints and 34 free parameters (42 when including unknown redshifts).

The parameters of the best-fit model (using the final catalogue of all
multiply imaged systems) are listed in Table~\ref{tab:model}. The
ellipticities of the two main components are given as the ratio of
minor/major axis ($b/a$) and the position angle $\phi$ measured
west-through-north. The parametrised model predicts the images with an
average accuracy of $3\arcsec$. The least well predicted image system
is the image system F, which is however perturbed by one of the
cluster members (included in the model only as one of the 20 cluster
members and not minimized for individually). The two main components have
very similar velocity dispersions, however as discussed extensively in
\citet{halkola08} given the degeneracies in the modelling, this is not
the only solution for this cluster.

We justify the choice of a two component model by comparing the
Bayesian evidence (for details see \citealp{kneib03,jullo07}) when
fitting one or two PIEMD components (both times including the cluster
members to the fit, and letting all the parameters of one or both
PIEMD components to vary). The logarithmic evidence is a factor of 3
larger in the case of a two component model, thereby clearly
justifying the two-clump model used for the fit.

Using the samples (i.e. a set of models that satisfactory fits the
data) we then predict possible additional images for the initial guess
system and matches to further (not necessarily yet identified as
multiple) images. This allows us to find a region in space, where
matches to known systems and objects should be searched for. Further,
if a match in color space is found, the hypothesis that it is a
multiply imaged system can be tested and the redshift estimate is
obtained using the improved lens model.

All the identified images were visually inspected (we are hoping to be
able to omit this step altogether in the future).

\subsection{Redshifts for Strongly Lensed Systems}

\begin{figure*}[ht]
\begin{center}
\includegraphics[width=0.6\textwidth]{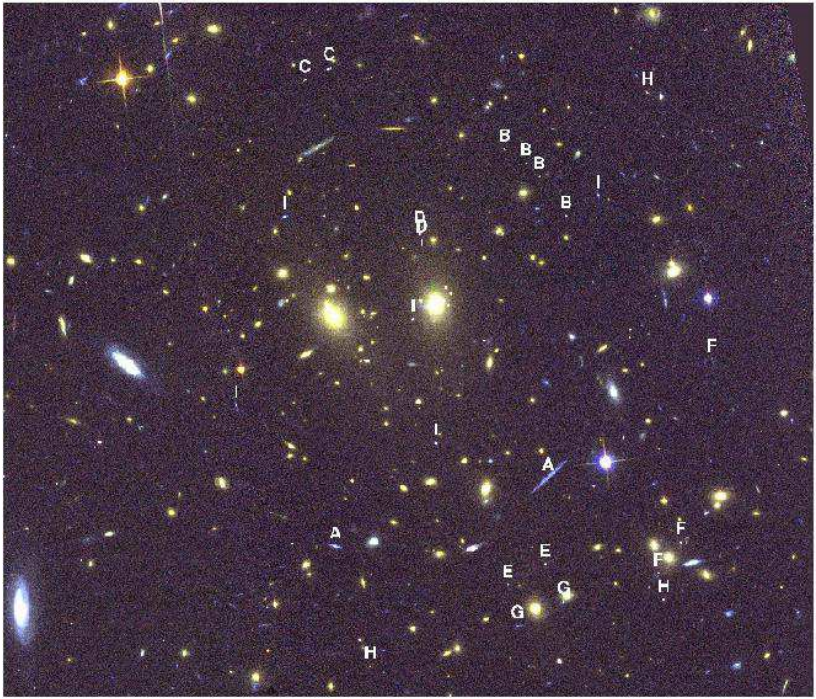}
\end{center}
\caption{The F450W-F814W-F850LP color composite of the cluster
  \protect \rxj. Multiply imaged systems are marked and labeled (see also \protect{Table~\ref{tab:arcs}}).  North is up and East is left, the field is
  $2.3^{\prime}\times 2.3^{\prime}$, which corresponds to $800 \times
  800 \mbox{ kpc}^2$ at the redshift of the cluster. The color
  composite was created following the algorithm from \citet{lupton04}.}
\label{fig:arcs}
\end{figure*}

For {\rxj} the only multiply imaged system with confirmed
spectroscopic redshift is system A (Table~\ref{tab:arcs} and
Fig.~\ref{fig:arcs}).  The redshift of the brighter of the two images
was independently measured to be 1.75 for both images with FORS
spectroscopic data \citep{lombardi07} as well as Keck data. This is in
agreement with the photometric redshift obtained from UBVRIJHKs
ground-based data in
\citet{bradac04b}, where a redshift $1.76 \pm 0.1$ was used for this system. We
have predicted the redshifts of the other identified systems by
setting them as free parameters when the system was added in the
image identification procedure outlined above. The resulting best-fit
redshifts are given in Table~\ref{tab:arcs}. The final error bars
were obtained by source-plane modelling (modelling in the image-plane
with all the parameters set free would be too time consuming) and
should be treated as approximate.

Ideally, the unknown redshift should be fitted simultaneously with
weak and strong lensing data using the method described in
\S\ref{sec:resultsw} (which does not assume a specific model family for
the cluster potential). This is unfortunately not possible due to the
so-called mass-sheet degeneracy \citep{falco85,seitz95}. This
transformation essentially allows us to determine the surface mass
density only up to a constant (that enters in a re-scaling and offset
of the surface mass density). Since the redshift information enters
the lensing equation in similar fashion (see
\citealp{bradac03} for more details), if only a {\it single} redshift
of multiply imaged system is known, without assuming a parametrized
model, the rest of the redshifts can not be determined. Weak
lensing in principle should provide a second redshift plane and hence
constrain the profile and its normalisation, allowing us to reconstruct
the data without the need of any simple parametrised modelling. In
practice, however, the weak lensing data is noisy and therefore not
sufficient to break the mass sheet degeneracy (see
e.g. \citealp{bradac03, limousin07}). If one tries to constrain the
unknown redshift of a system with images far away from the images with
known redshift, one can obtain equally good fits by either slightly
perturbing the potential or by changing the redshift. Hence we need to
use parametrized models for this part of the analysis. However, when
other multiply imaged systems are present in the vicinity, the
redshift can still be determined using the pixelated model;
when we added image system I we could only get a good fit with $z=1.7$
and not the photometric redshift estimate of $z=2.2$; the same conclusion
was obtained already by \citealp{halkola08}.

\begin{deluxetable}{rrrr}
\tablecolumns{4}
\tablewidth{0pc}
\tablecaption{The properties of the multiply-imaged systems used in this work. The redshift of the system A was obtained using FORS spectra, whereas the others have been determined by leaving them as free parameters when fitting strong lensed data with parametrised model.}
\tablehead{ \colhead{} &  \colhead{Ra} & \colhead{Dec} & \colhead{$z_{\rm pred}$}}
\startdata
& 206.87207 & $-$11.761072 & \\
\raisebox{0.5ex}{A} & 206.88263 & $-$11.764407 & \raisebox{0.5ex}{1.75} \\ 
\cline{1-4} 
 & 206.87250 & $-$11.746470 &  \\ 
 & 206.87117 & $-$11.748356 &  \\ 
\raisebox{0.5ex}{B} & 206.87316 & $-$11.745810 & \raisebox{0.5ex}{$1.2 \pm 0.1$} \\ 
 & 206.87423 & $-$11.745110 &  \\ 
\cline{1-4} 
 & 206.88293 & $-$11.741210 & \\ 
\raisebox{0.5ex}{C} & 206.88415 & $-$11.741798 & \raisebox{0.5ex}{$2.0\pm 1.0$} \\ 
\cline{1-4} 
 & 206.87845 & $-$11.749200 & \\ 
\raisebox{0.5ex}{D} & 206.87833 & $-$11.749682 & \raisebox{0.5ex}{$2.2\pm 0.1$} \\ 
\cline{1-4} 
 & 206.87219 & $-$11.765260 &  \\ 
\raisebox{0.5ex}{E} & 206.87406 & $-$11.766258 & \raisebox{0.5ex}{$2.5 \pm 0.6$} \\ 
\cline{1-4} 
 & 206.86549 & $-$11.764203 &  \\ 
{F} & 206.86661 & $-$11.765704 &{$4.0\pm 2.0$} \\ 
 & 206.86395 & $-$11.755330 &  \\ 
\cline{1-4} 
 & 206.87357 & $-$11.768255 & \\ 
\raisebox{0.5ex}{G} & 206.87130 & $-$11.767085 & \raisebox{0.5ex}{$3.0\pm 0.5$} \\ 
\cline{1-4} 
 & 206.88088 & $-$11.770223 &  \\ 
{H} & 206.86634 & $-$11.766999 & {$4.2\pm 1.0$} \\ 
 & 206.86714 & $-$11.742389 & \\
\cline{1-4}
 & 206.88754 & $-$11.757578 &  \\
 & 206.88513 & $-$11.748397 &  \\
I & 206.87878 & $-$11.753372 & {$1.7\pm 0.2$} \\
 & 206.87765 & $-$11.759400 &  \\
 & 206.86960 & $-$11.747372 & \\
\cline{1-4}
\enddata
\label{tab:arcs}
\end{deluxetable}

\begin{deluxetable}{rrrrrr}
\tablecolumns{6} \tablewidth{0pc}
\tablecaption{Results of best-fit parametric modelling of strong lensing data (parameters refer to the values used by {\tt LENSTOOL}, axis-ratio $b/a$ as that of the projected mass distribution).}
\tablehead{\colhead{} & \colhead{$\sigma_{\rm PIEMD}$}& \colhead{$r_{\rm c}$}& \colhead{$r_{\rm cut}$} & $b/a$ & \colhead{$\phi$} \\ \colhead{} & \colhead{$(\mbox{ km}\,\mbox{s}^{-1}$)} & \colhead{$(\mbox{kpc})$} & \colhead{$(\mbox{kpc})$} & \colhead{} & \colhead{} }
\startdata
BCG & $1640 \pm 40$ & $340 \pm 30$& $1000\pm 30$ &$0.89 \pm 0.01$ & $16^{\circ} \pm 1^{\circ}$ \\ 
2nd cD & $1640 \pm 10$ & $400 \pm 5$ & $2000 \pm 60$ & $0.87 \pm 0.01$ & $70^{\circ} \pm 1^{\circ}$\\ 
\enddata
\label{tab:model}
\end{deluxetable}

\begin{figure*}[ht]
\begin{center}
\begin{tabular}{ccccc}
\includegraphics[width=0.18\textwidth]{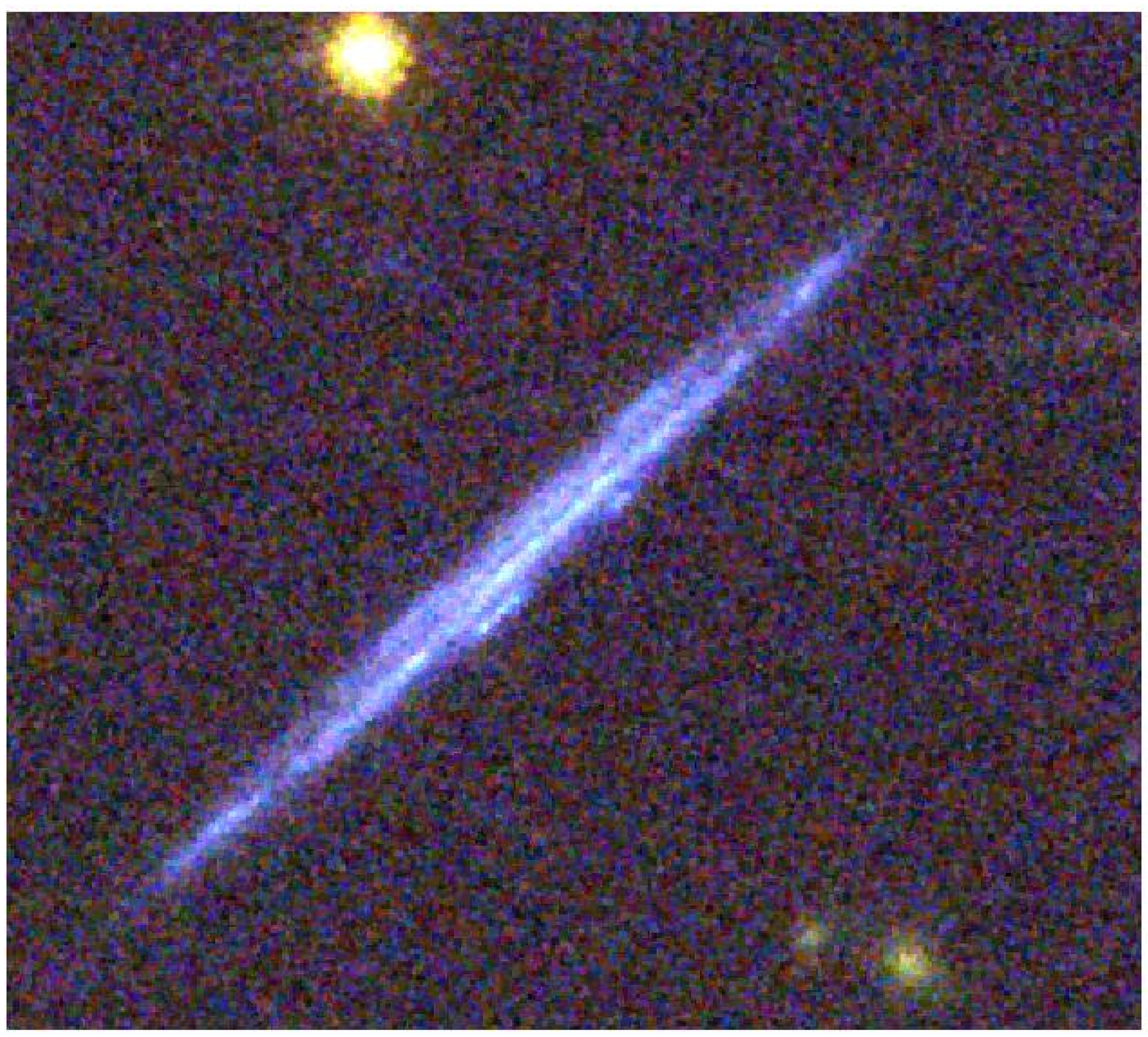} &
\includegraphics[width=0.18\textwidth]{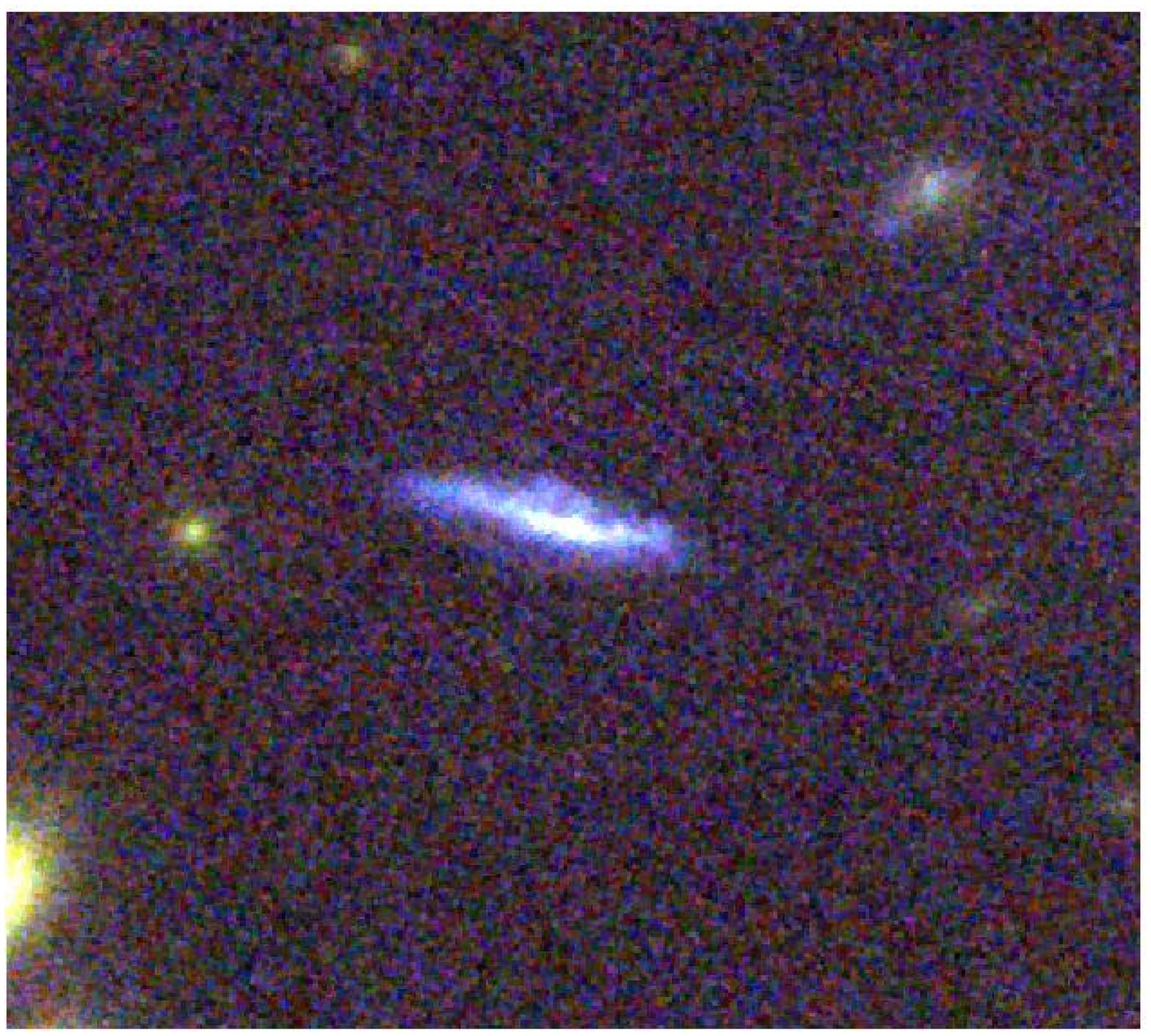} &
\includegraphics[width=0.18\textwidth]{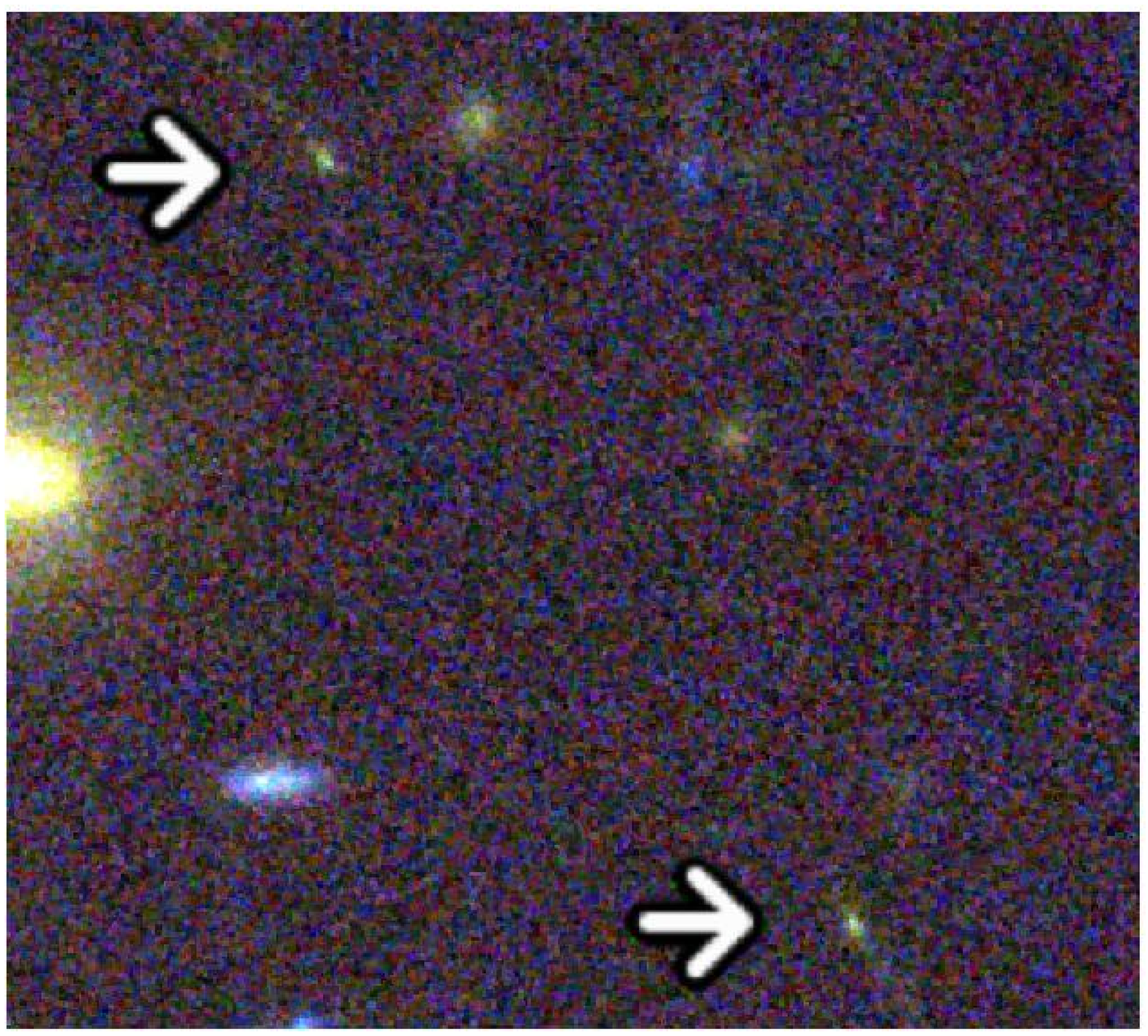} &
\includegraphics[width=0.18\textwidth]{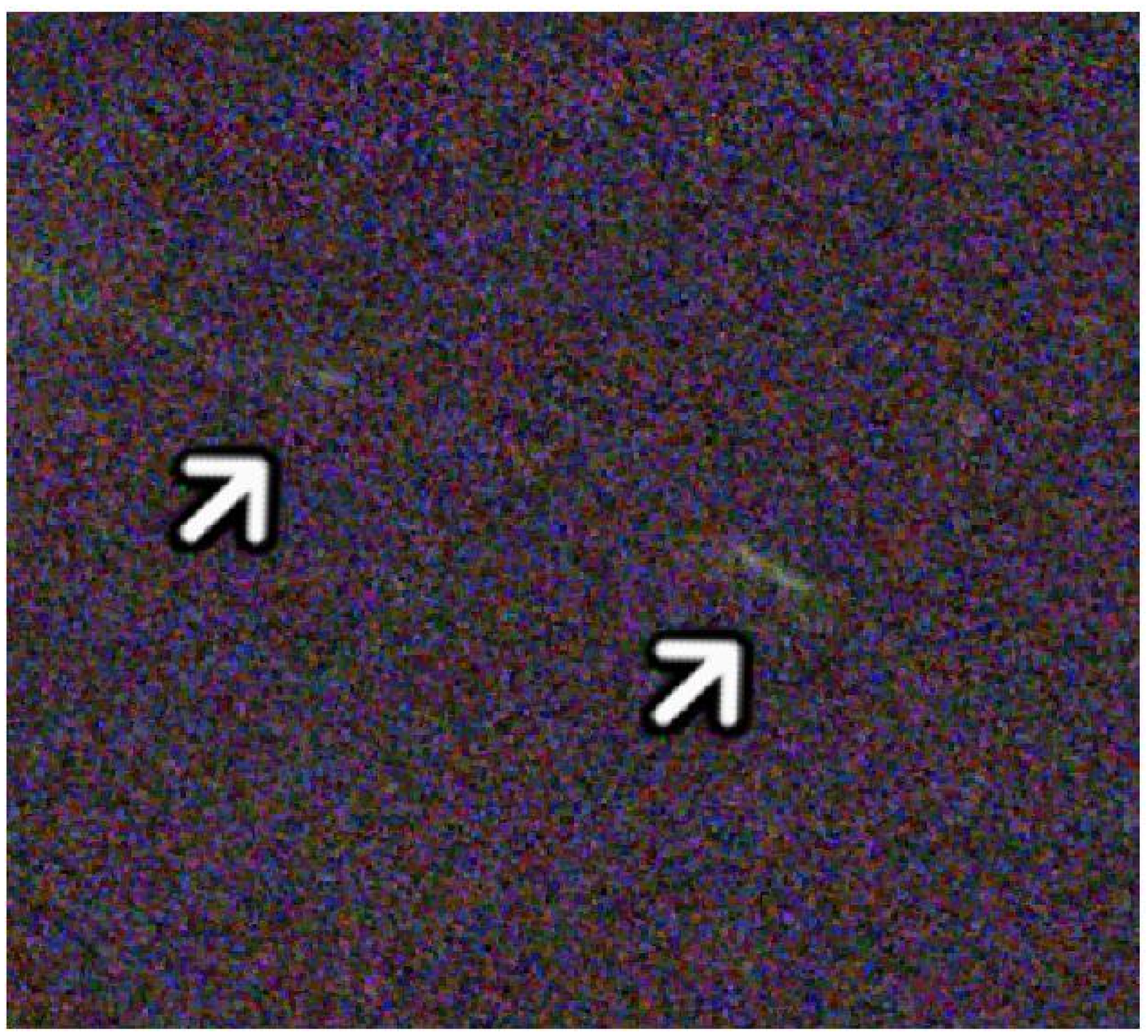} & 
\includegraphics[width=0.18\textwidth]{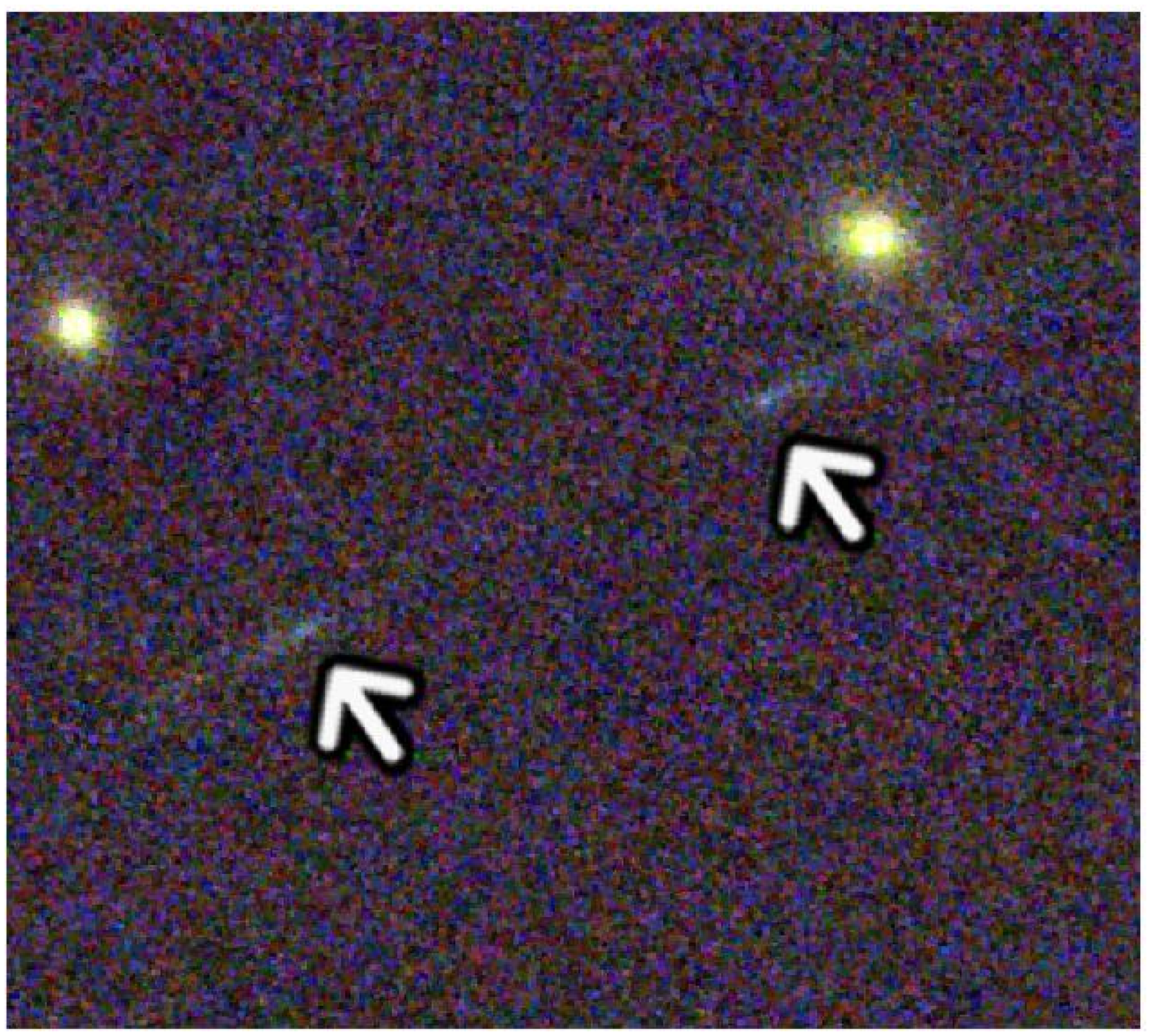} \\
A & A & B & B & C \\
\includegraphics[width=0.18\textwidth]{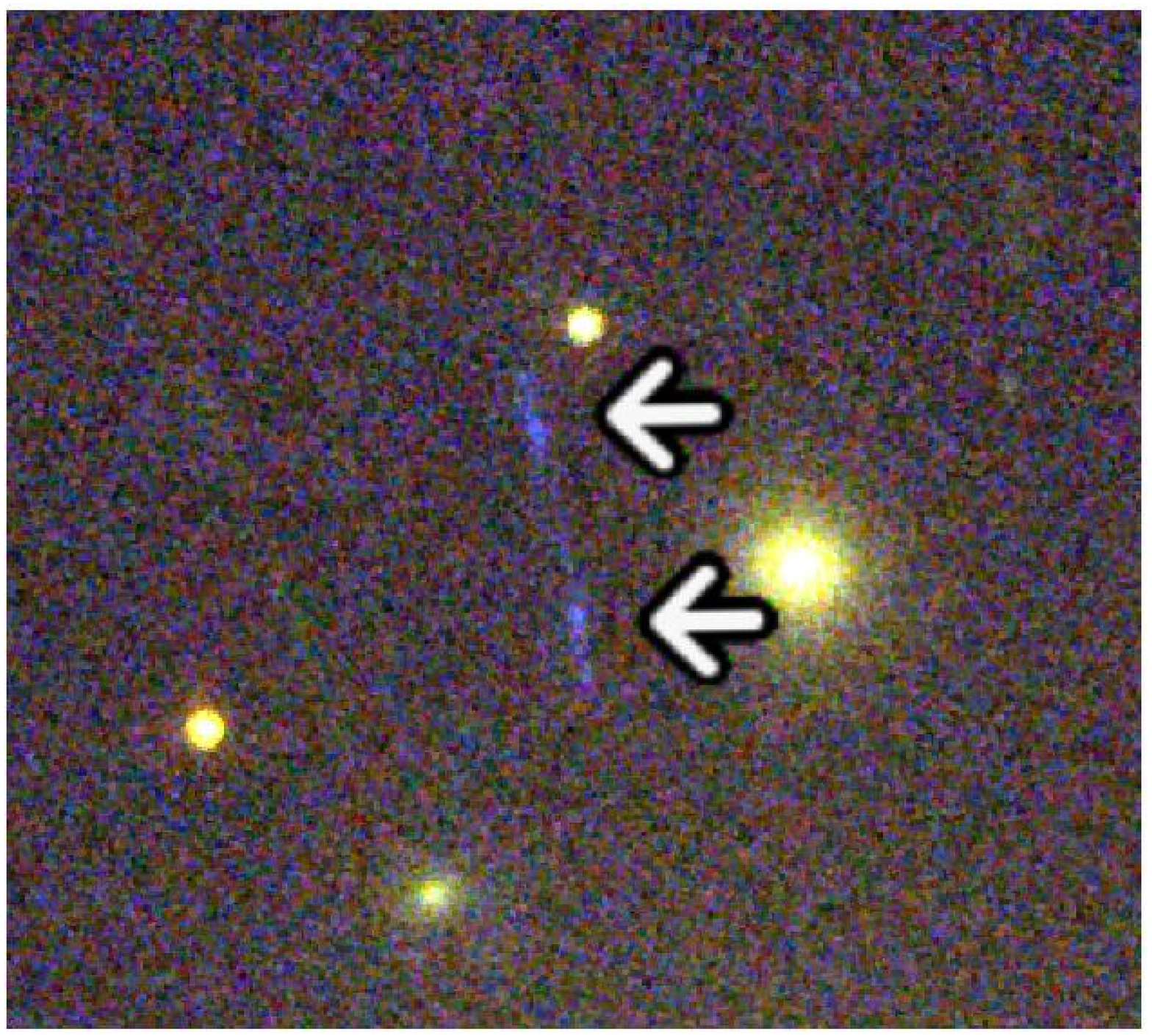} &
\includegraphics[width=0.18\textwidth]{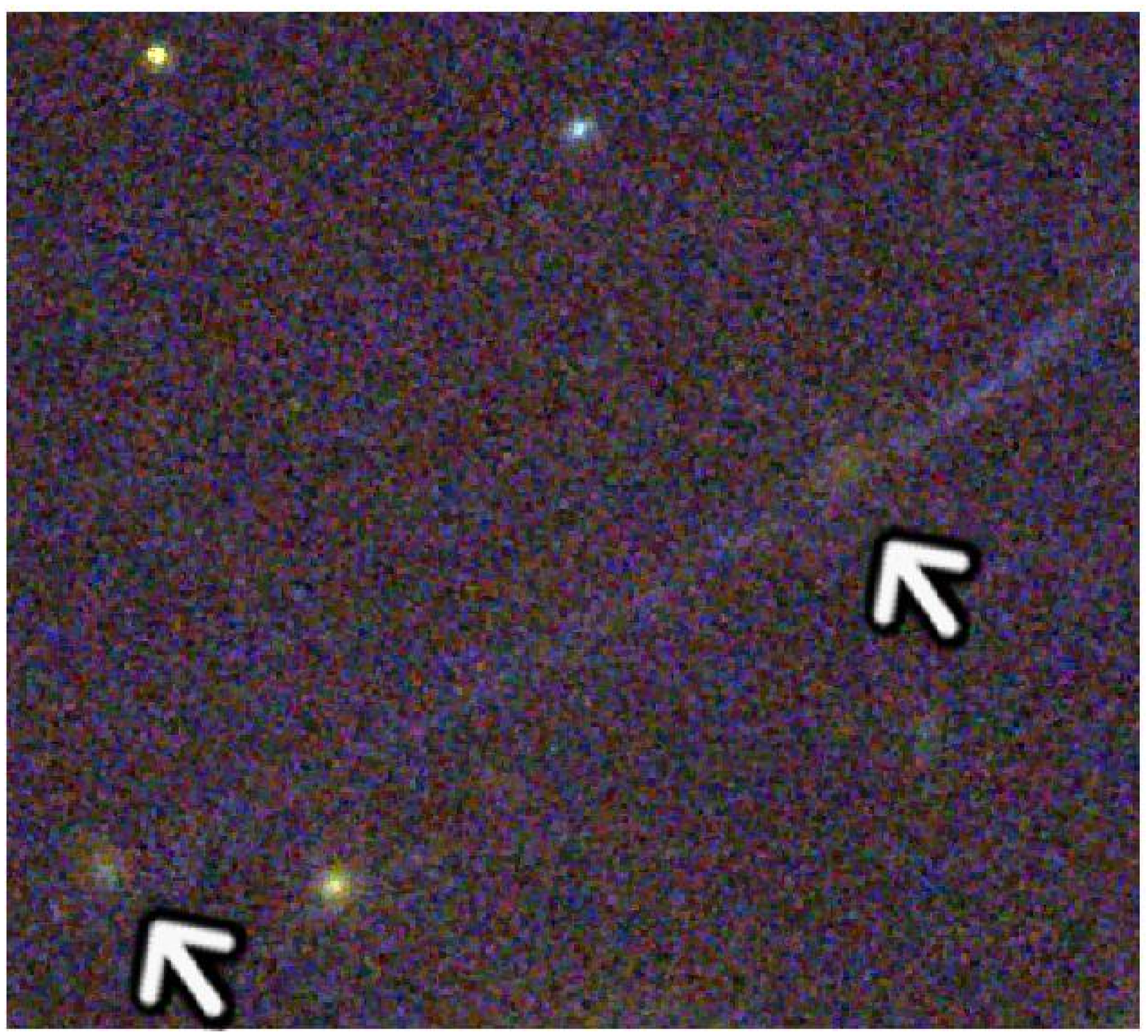} &
\includegraphics[width=0.18\textwidth]{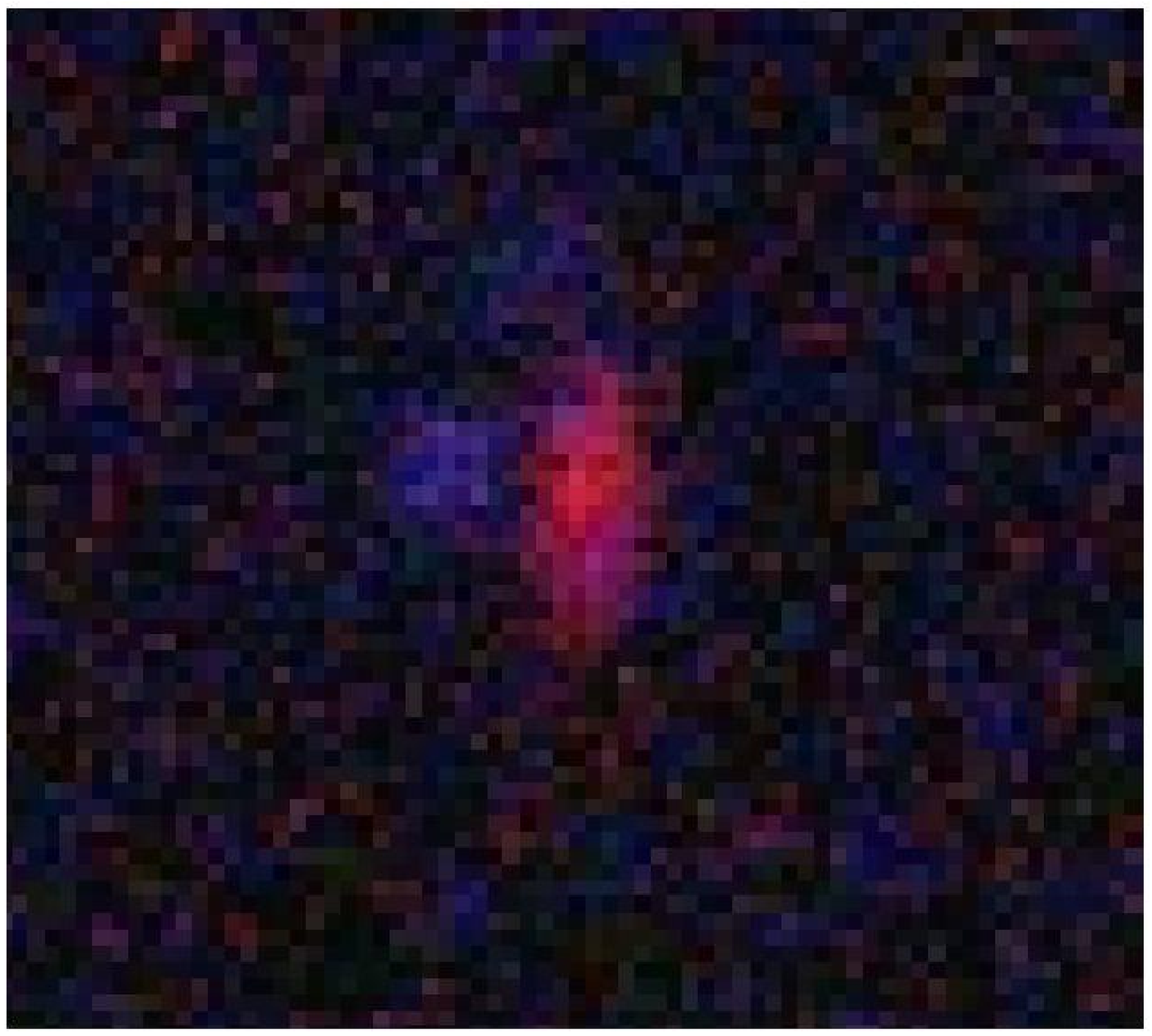} &
\includegraphics[width=0.18\textwidth]{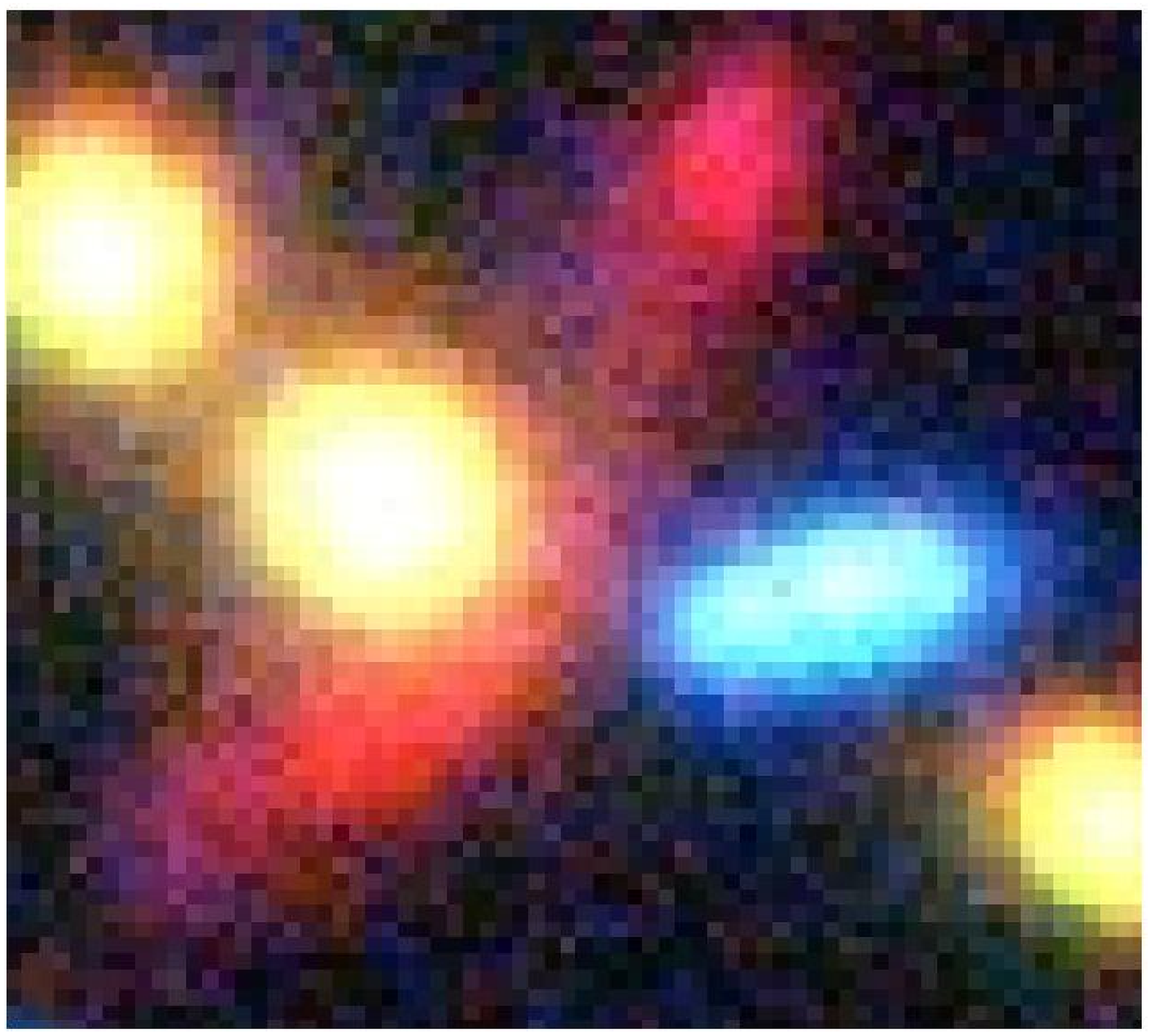} &
\includegraphics[width=0.18\textwidth]{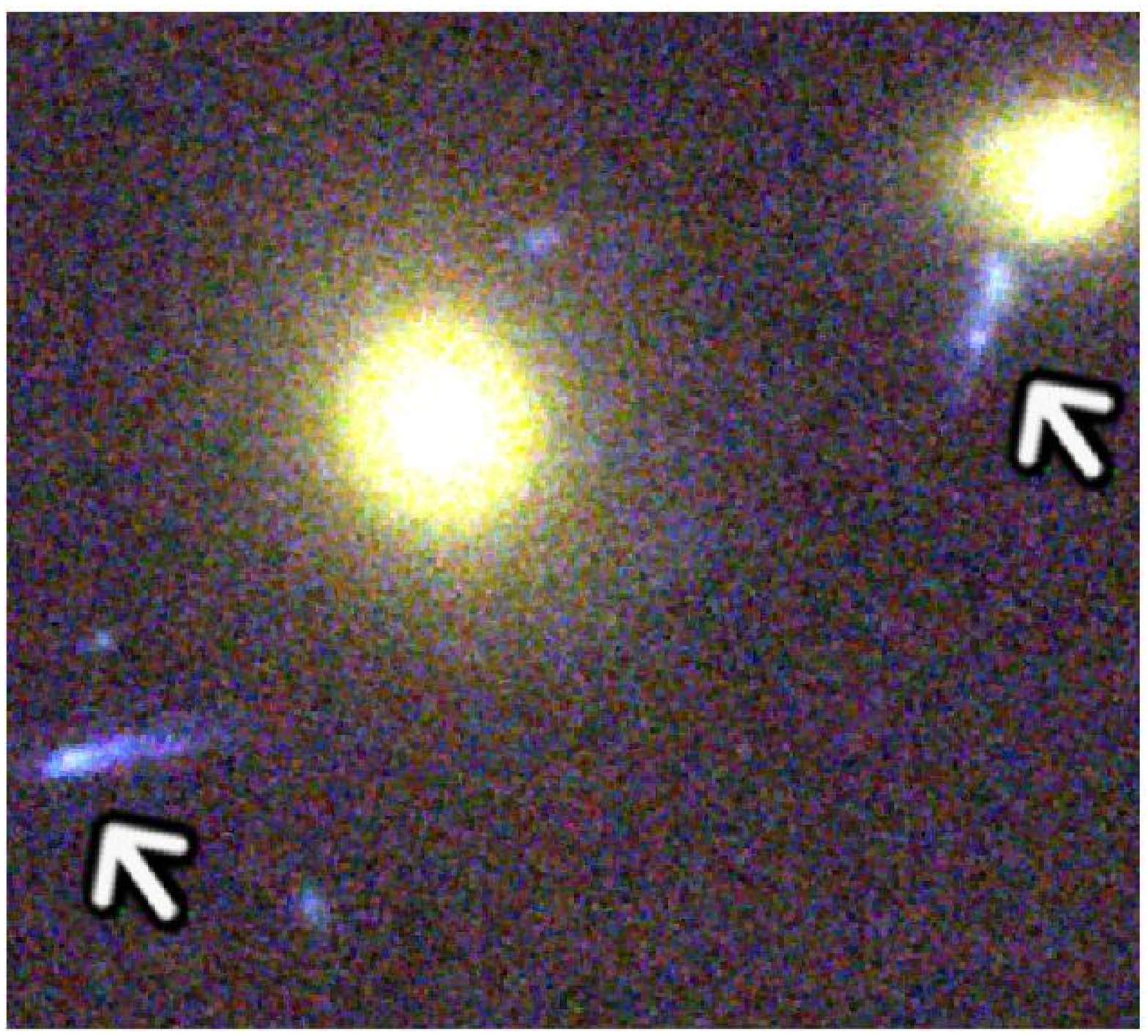} \\
D &E & F & F & G \\
\includegraphics[width=0.18\textwidth]{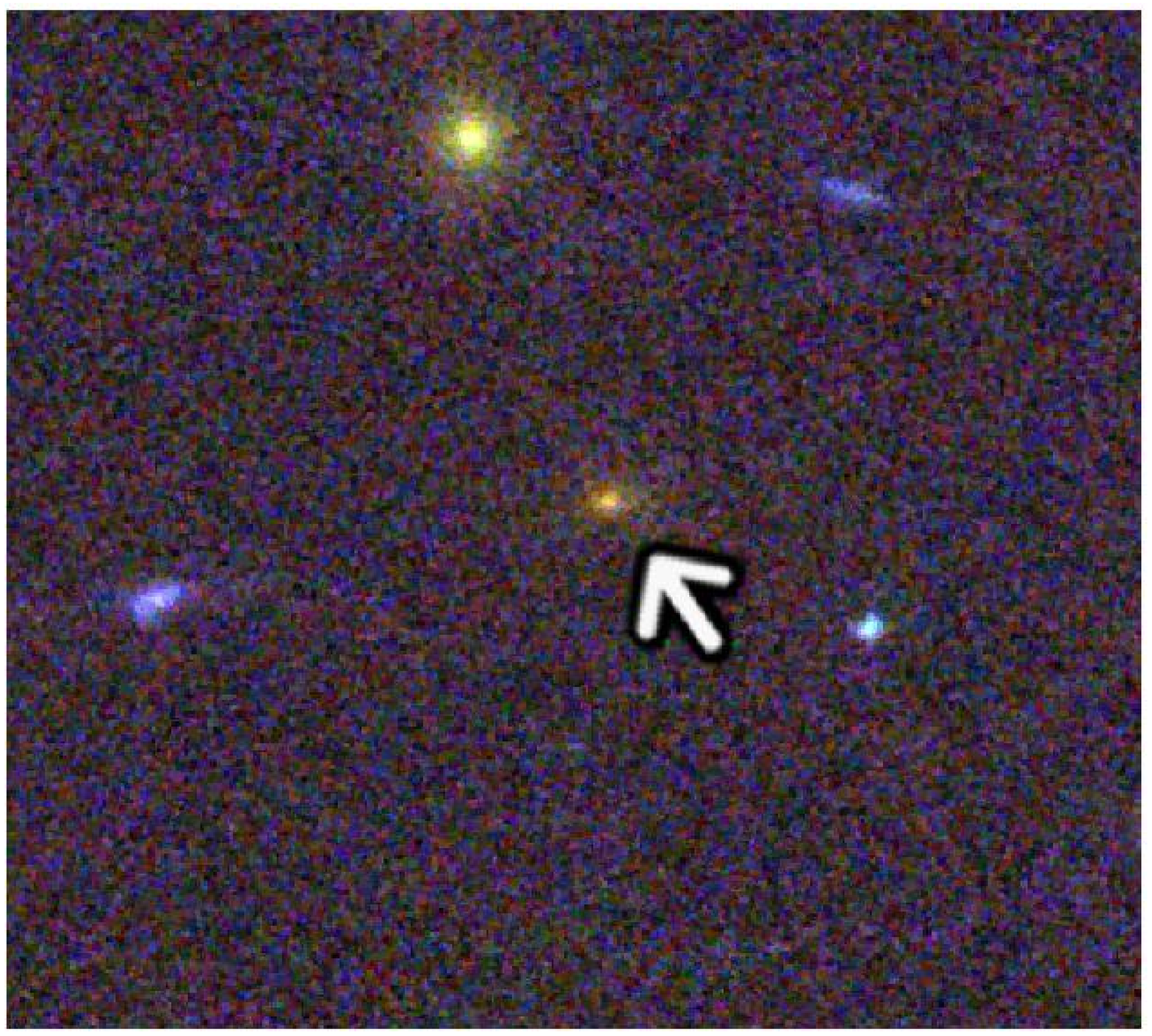} &
\includegraphics[width=0.18\textwidth]{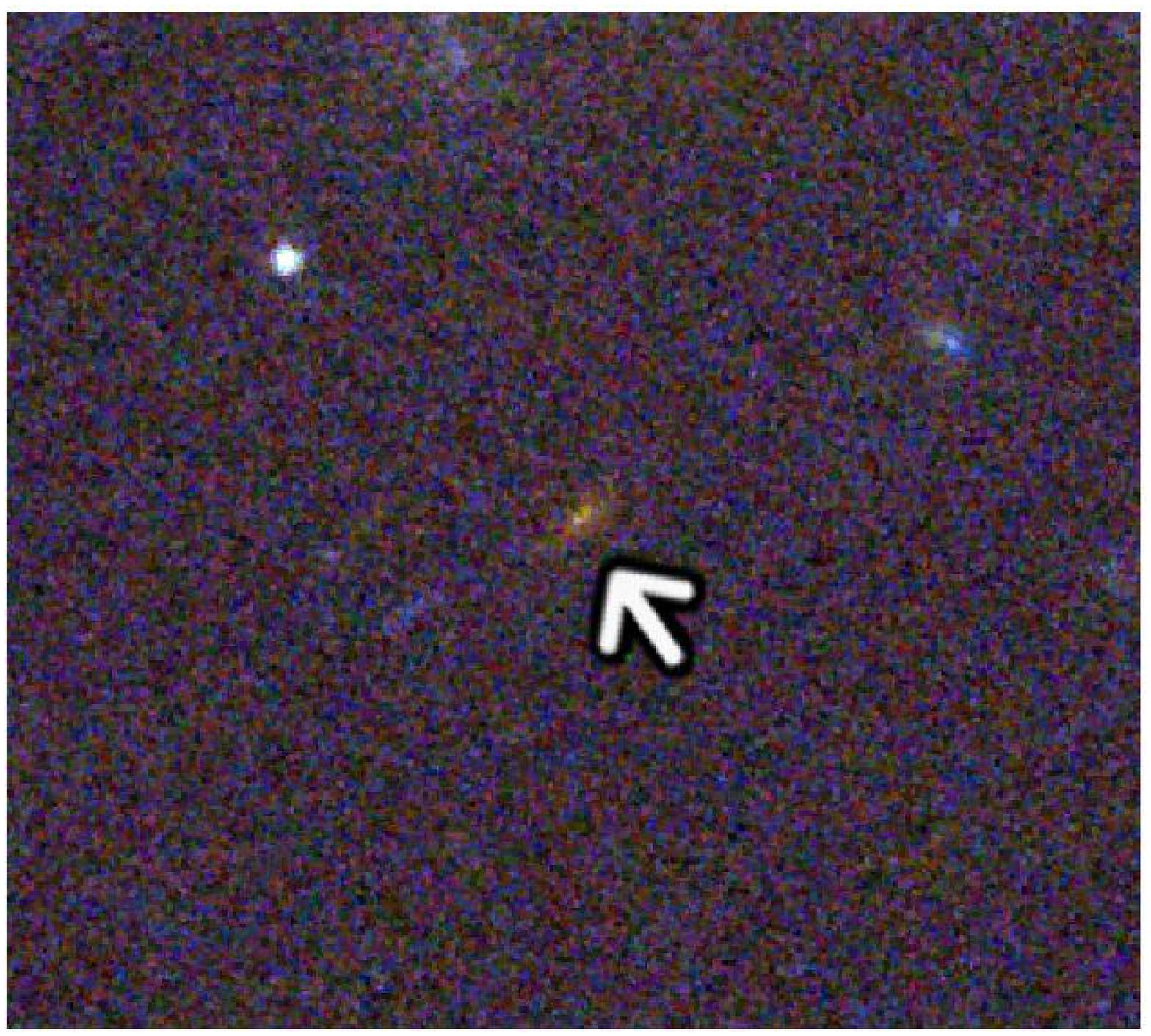} &
\includegraphics[width=0.18\textwidth]{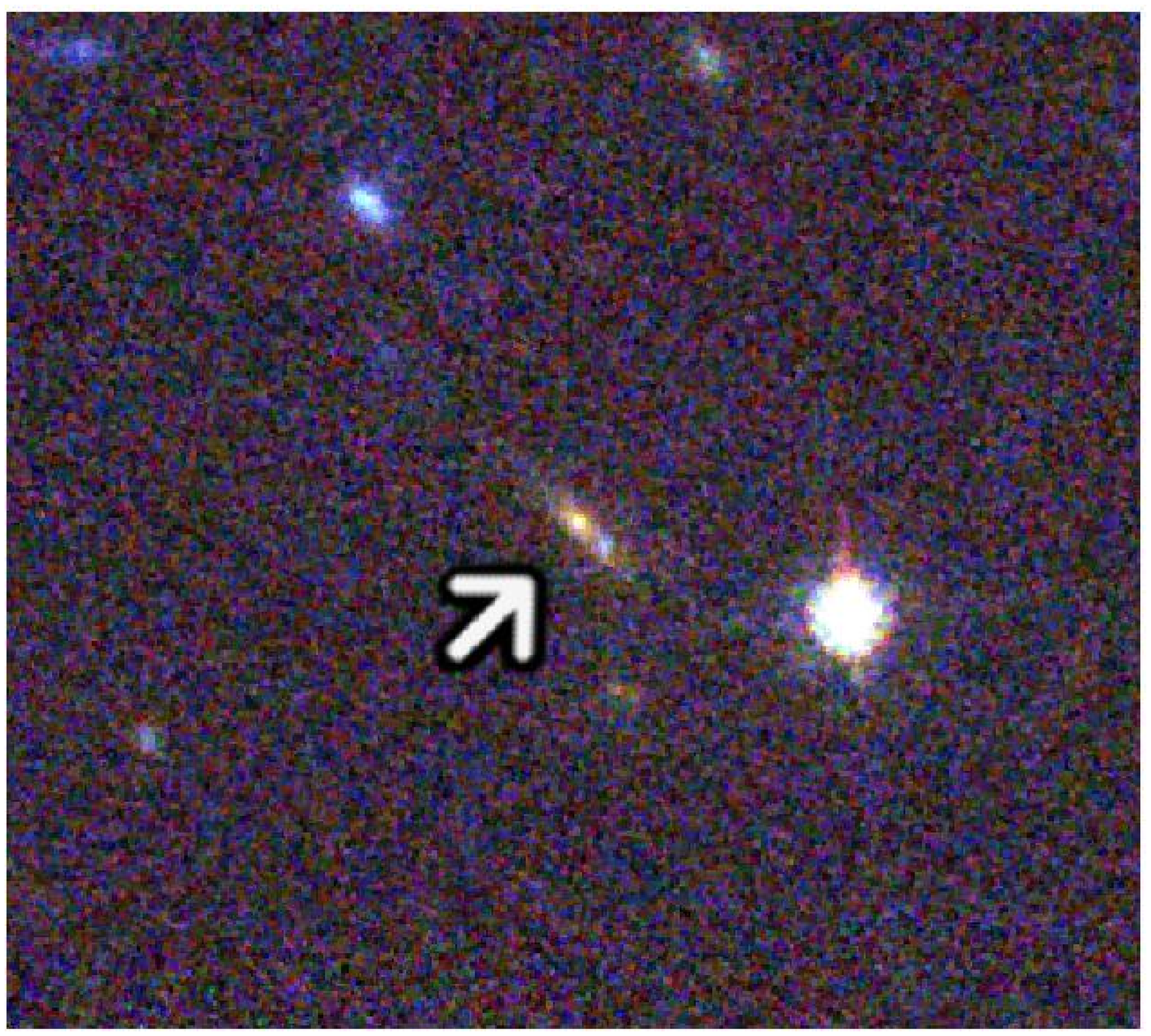} \\
H & H & H & &\\
\includegraphics[width=0.18\textwidth]{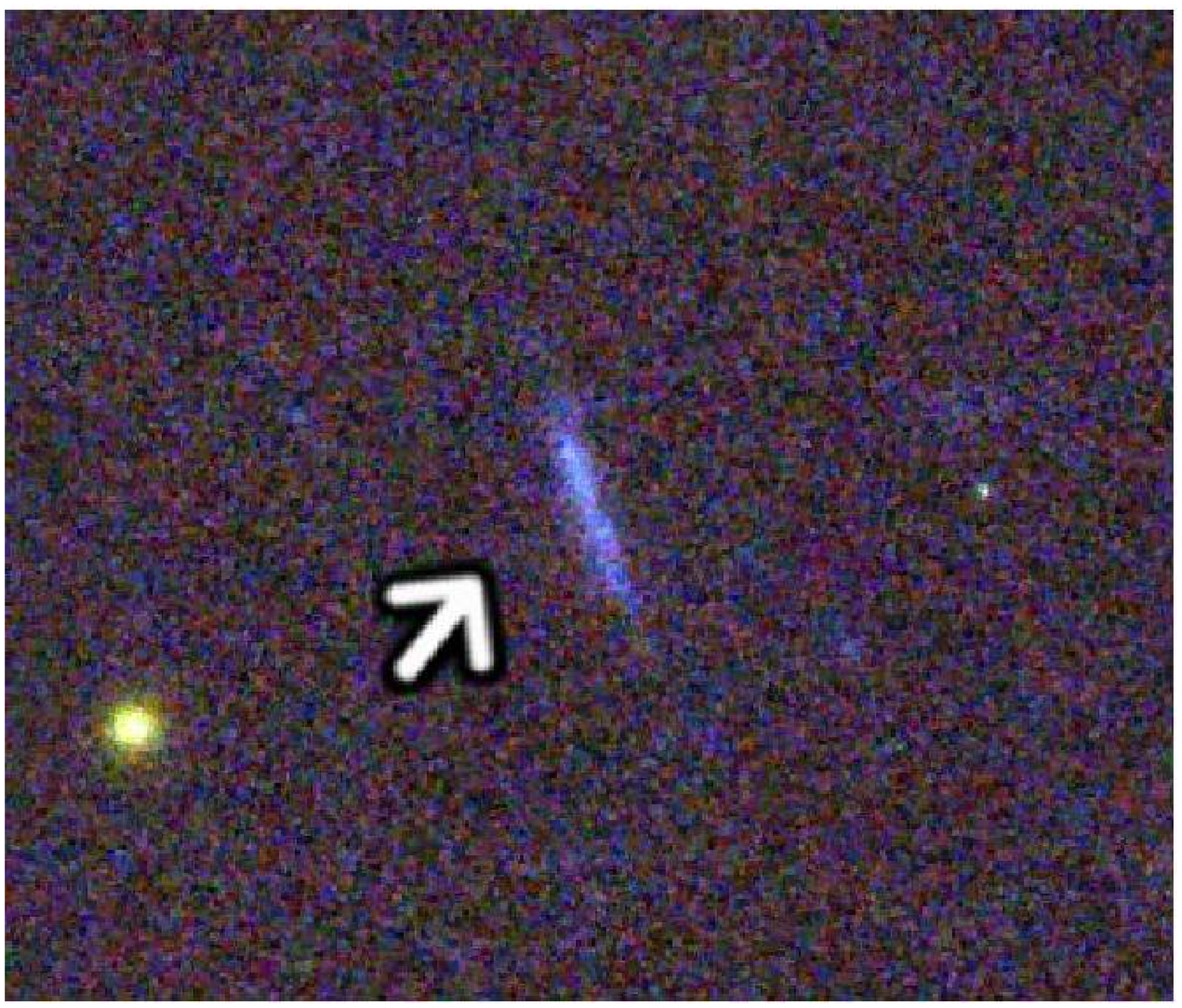} &
\includegraphics[width=0.18\textwidth]{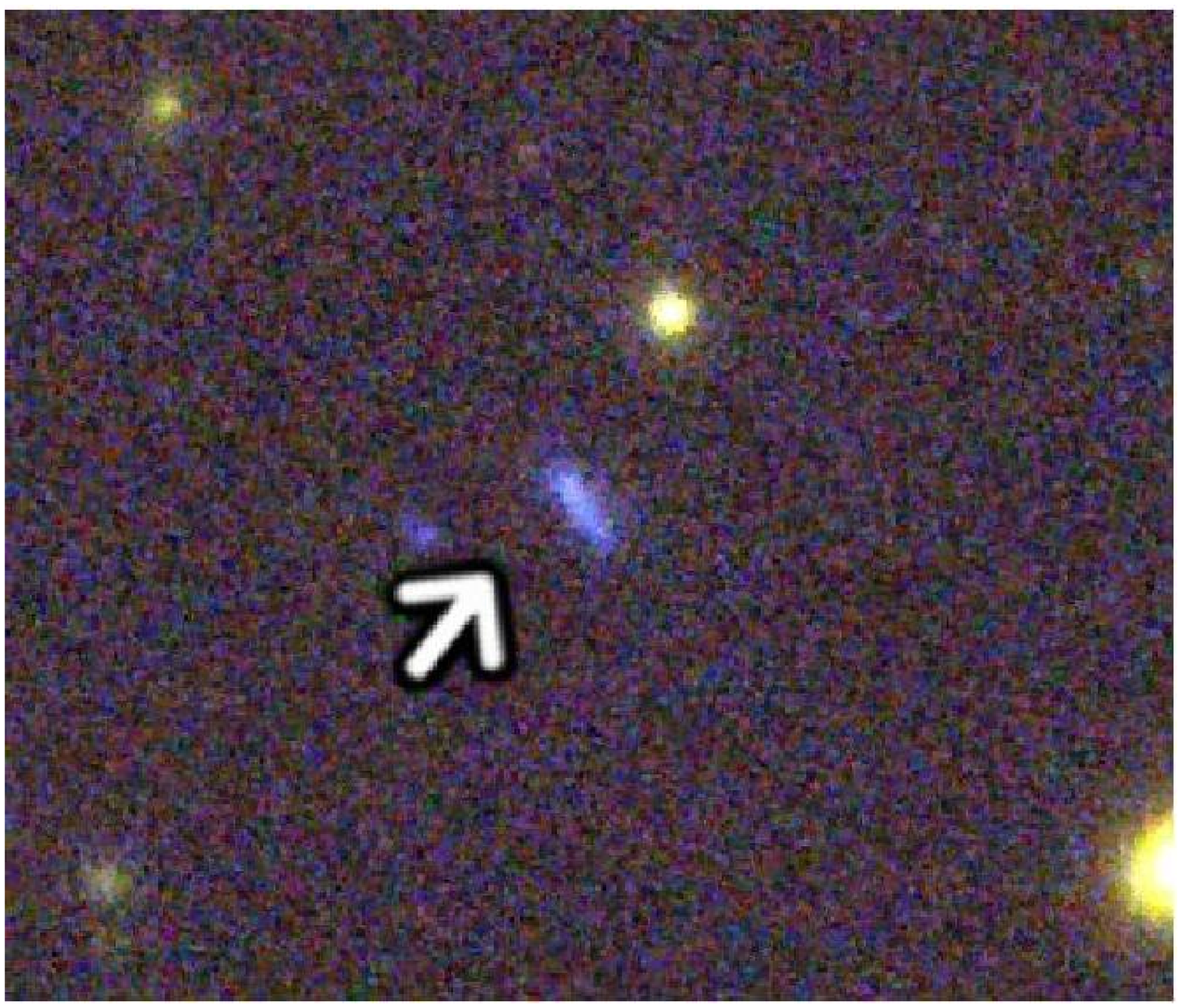} &
\includegraphics[width=0.18\textwidth]{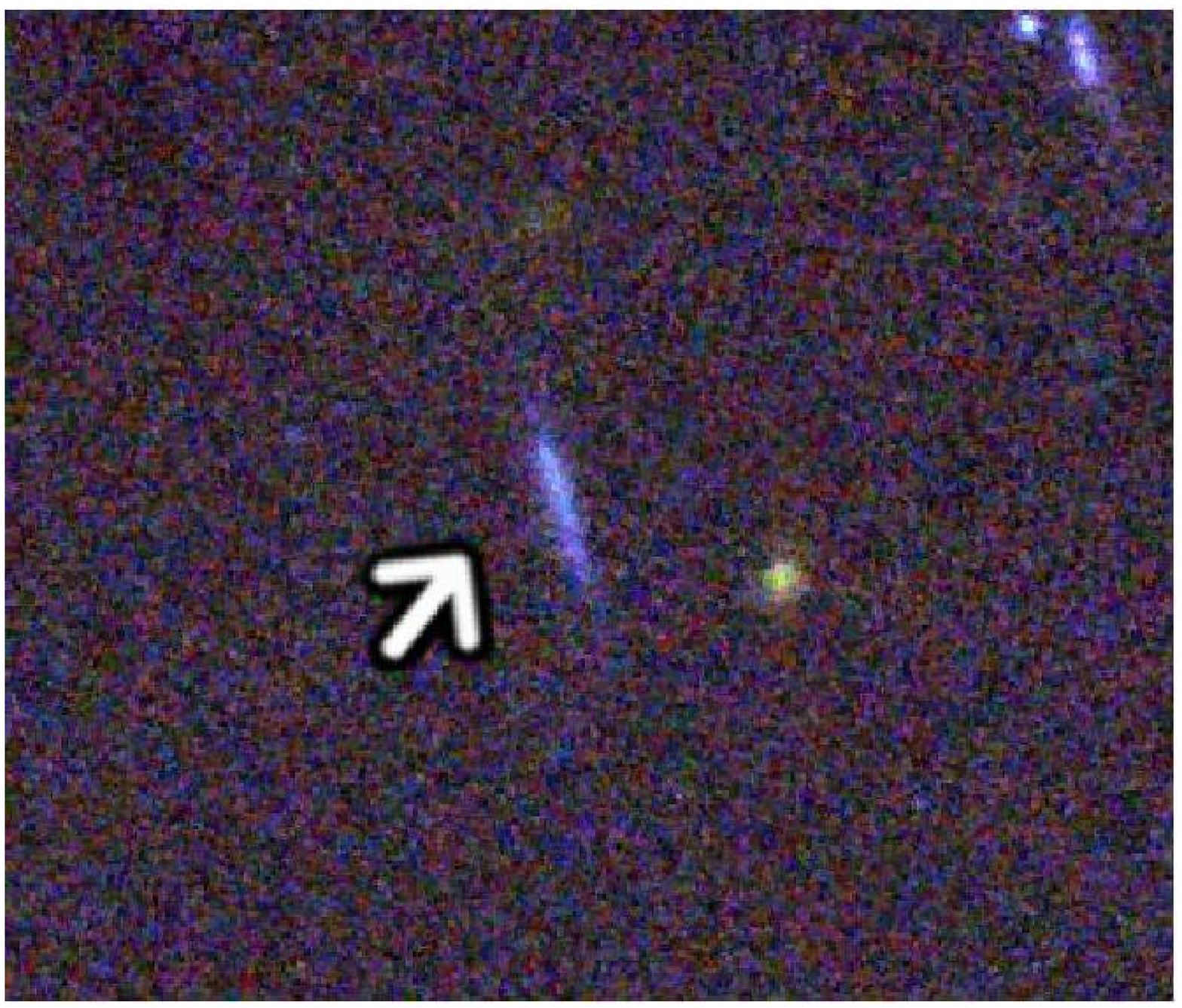} &
\includegraphics[width=0.18\textwidth]{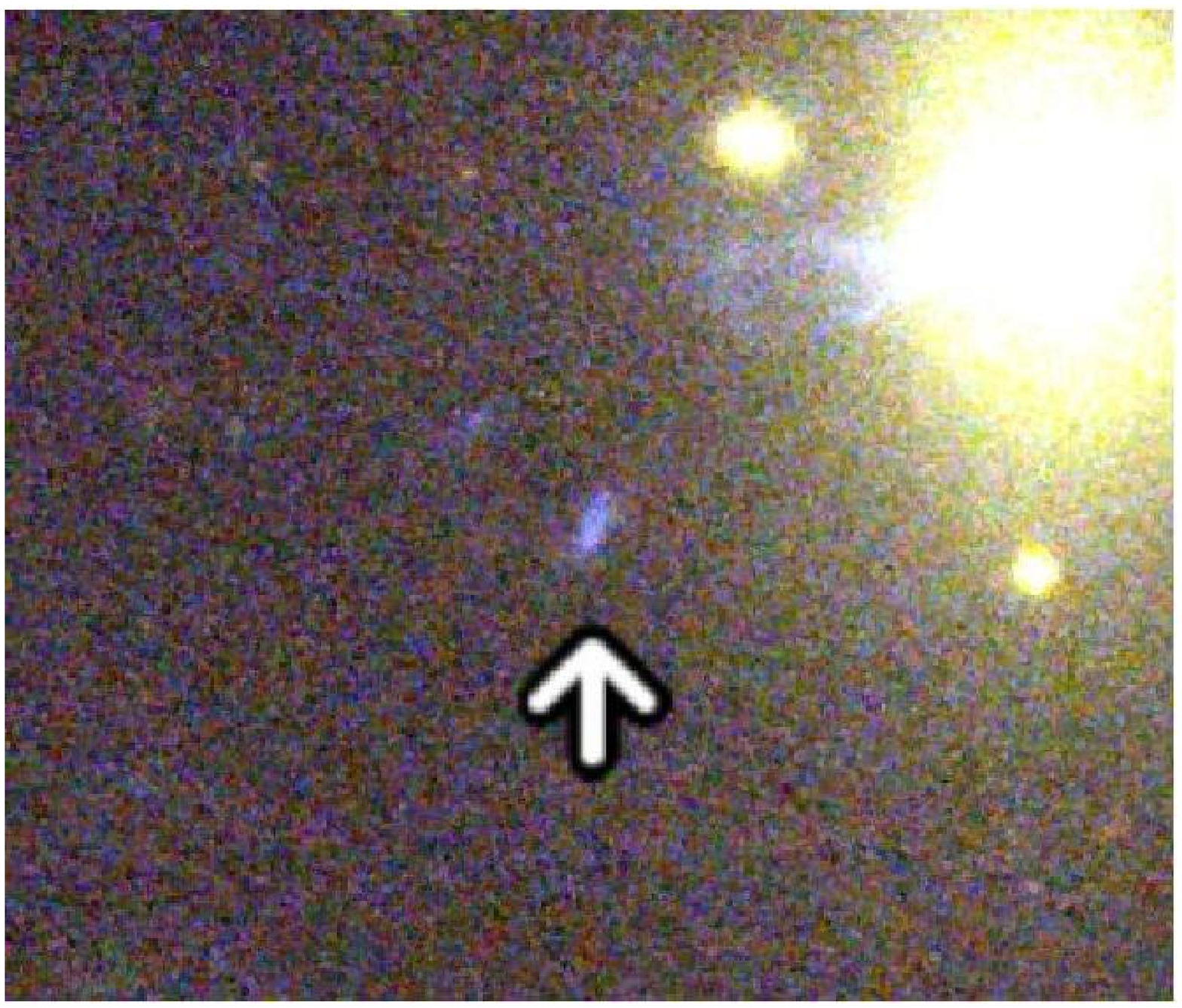} &
\includegraphics[width=0.18\textwidth]{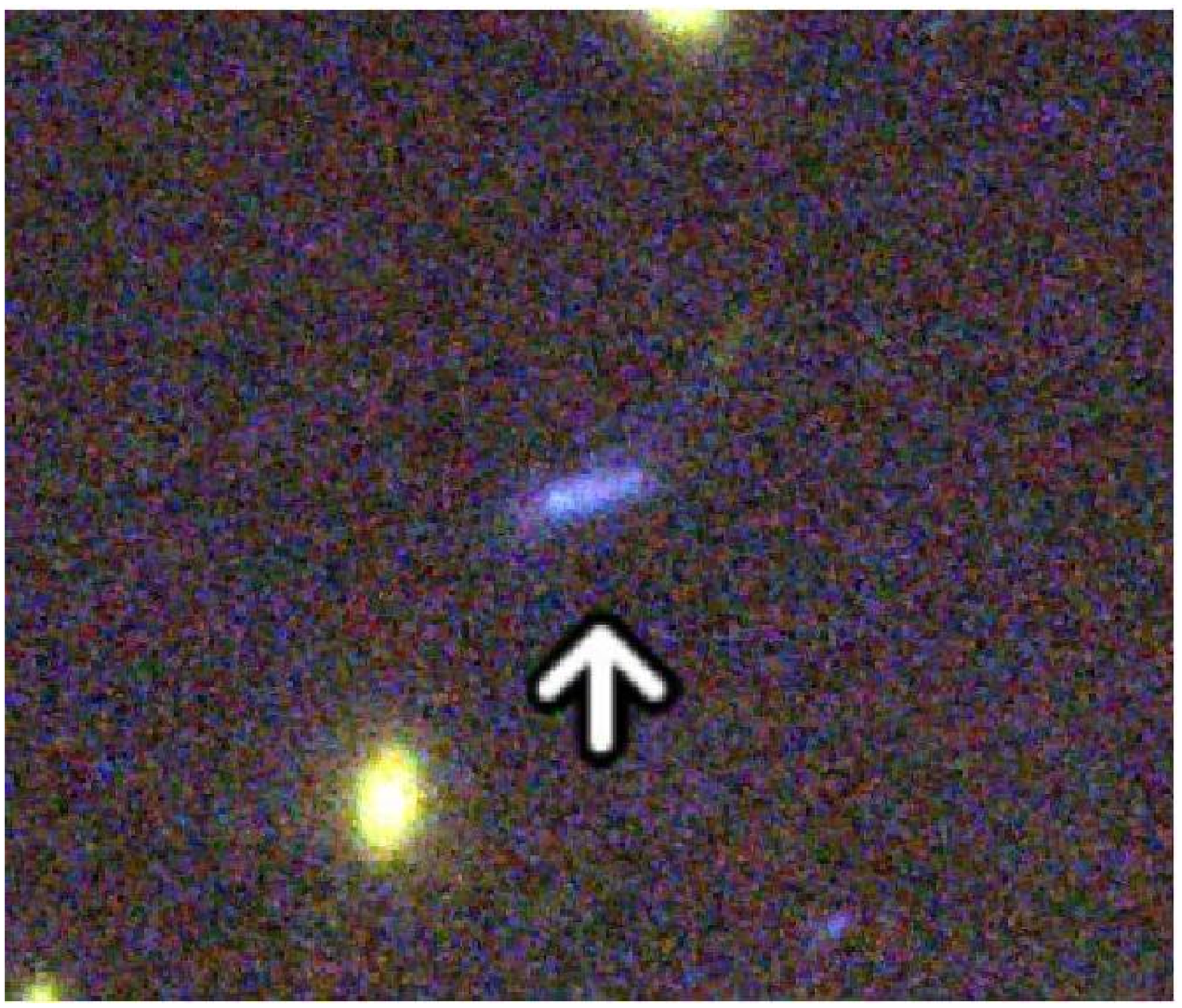} \\
I & I & I & I & I\\
\end{tabular}
\end{center}
\caption{$10\arcsec \times 10\arcsec$ cutouts of the multiply imaged systems from the F450W-F814W-F850LP color composite (except for image F, where BRK FORS data was used due to its extreme red color).}
\label{fig:arcs2}
\end{figure*}

\section{Weak Lensing Data Analysis}
 \label{sec:wlensing}

We perform the weak lensing shape measurements
on the F814W data due to their higher galaxy number density.  For the
PSF anisotropy and smearing correction we closely follow the technique
described in \citet{schrabback07} The procedure is based on the KSB+
algorithm \citep{ksb95, luppino97, hoekstra98}, which has extensively
been tested on the simulations of the STEP collaboration
\citep{heymans05,massey07}. The KSB+ method is formally valid only in
the weak lensing regime, however as discussed in \citet{bradac06} (see
also \citealp{massey07b}) the approach is also valid in the non-weak
lensing regime to the accuracy needed here. We select stars from the
half-light-radius vs. magnitude diagram.

In contrast to \citet{bradac06} we use the new PSF correction scheme
developed by \citet{schrabback07}. The method is superior, since it
takes the temporal (as well as the spatial) variability of the ACS PSF
into account. The PSF variation is measured from stars in individual
exposures (rather than in the co-added image as was traditionally
done). It is then compared with star-rich fields from archival
observations (stellar templates). The shapes of the stars in
individual exposures of our data set are first compared to the shapes
of stars from stellar templates and the best template is chosen. The
full PSF model is then composed of several templates; such a model
therefore captures the full spatial as well as temporal variability of
ACS. We detect short-term variations of the ACS PSF, which are
interpreted as focus changes due to thermal breathing of the telescope
(see also \citealp{schrabback07, krist03, rhodes07, anderson06}). The detailed
PSF model obtained from this procedure is then used to correct the
shapes of all the objects in the {\rxj} field.

Finally, since the PSF shape of stars changes as a function of the
scale at which we use for the shape measurement (see
e.g. \citealp{heymans05,jee06}), we match the scale used for PSF and
galaxy shape measurement. This method reduces the systematic
contribution to the shear correlation functions due to PSF distortions
to $ < 2 \times 10^{-6}$ for galaxy fields containing at least 10
stars, and was demonstrated to be sufficient even for cosmic shear
measurements \citep{schrabback07}, where the signal is more than an
order of magnitude smaller than measured in cluster fields. In
addition, we apply a parametric model, similarly to the one proposed
by \citet{rhodes07}, to correct for the influence of the degraded
charge-transfer-efficiency on galaxy shapes, see Schrabback et al. (in
prep.) for further details.  Using simulated weak lensing data from
the STEP1 project \citep{heymans06}, we identified a constant bias
between the input and measured shear values. In the analysis of the
STEP2 data we correct for this bias by the introduction of a
multiplicative shear calibration factor of $1.10$, which prove to be
on average accurate at the $\sim 2\%$ level also for this different
set of image simulations \citep{massey07}.  In the analysis of the
{\rxj} data we apply slightly different galaxy selection criteria,
leading to a marginally changed calibration factor of $1.08$, which we
apply to the data.

To exclude cluster members from the final catalog we match it to the
photometric redshift catalog from the ground-based (VLT/FORS and
ISAAC) data from \citet{bradac04b}. We exclude all the galaxies having
measured photometric redshifts $z_{\rm phot} < 0.5$. Unfortunately,
however, we can not use the photometric redshift estimates for all the
object in the weak lensing catalog, since many of the faint objects
were undetected or unresolved in the ground-based data. Therefore
following the prescription of \citet{schrabback07} we use the
GOODS-MUSIC sample and apply the corresponding magnitude cuts to the
data and obtain the average redshift for galaxies in the weak
lensing catalogue of $z_{\rm WL} = 1.4$. The exact value we use has
little importance (as discussed in \S\ref{sec:syst}) on the final mass
reconstruction, since the weak lensing signal/noise is much lower than
the strong lensing one, and therefore the mass scaling is determined
predominately by the strong lensing systems. This is however only true
for the regions where we reconstruct the mass, once we extend the
measurements far outside the strong lensing regime with the ground
based data (beyond $\sim 500\mbox{ kpc}$) the determination of
redshift distribution will become much more important.

\section{X-ray Data Reduction}
 \label{sec:xraydata} 
\begin{figure}[ht]
\begin{center}
\begin{tabular}{c}
\includegraphics[width=0.5\textwidth, angle=270]{bradac_fig3a.ps}\\
\includegraphics[width=0.5\textwidth]{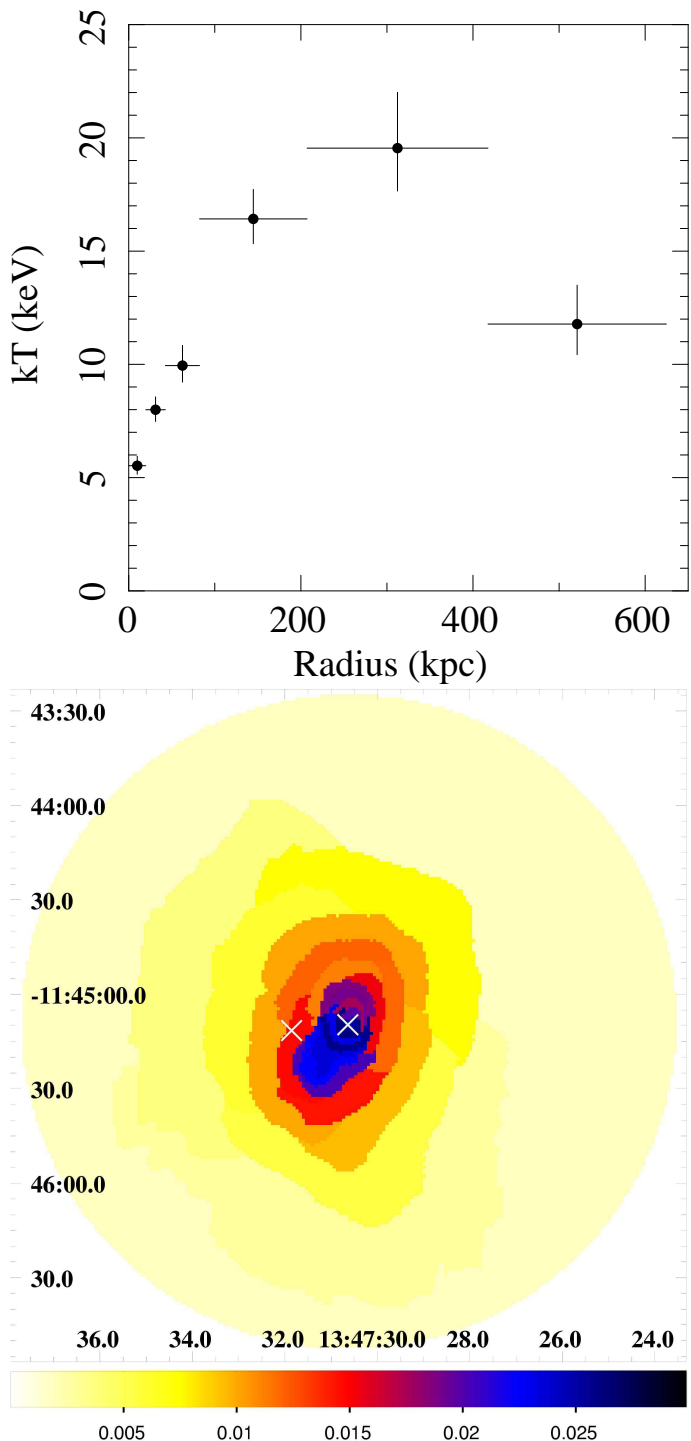}
\end{tabular}
\end{center}
\caption{(top) The deprojected temperature profile of the X-ray emitting gas in 
RXJ1347.5-1145 determined from the Chandra X-ray data. The data from the 
southeast quadrant were excluded from the analysis, as 
described in the text. (bottom) The pressure map of \protect \rxj. The two crosses indicate the two brightest cluster members.}
\label{fig:xray}
\end{figure}

Under the assumptions of hydrostatic equilibrium
and spherical symmetry, one can use the observed X-ray surface
brightness profile and the deprojected X-ray gas temperature profile
to determine the total mass and gas mass profile. The reduction and
analysis of the Chandra X-ray data is described in detail by
\citet{allen07}, \citet{schmidt07}, and Million et al. (in
preparation).  The thermodynamic X-ray pressure map determined from
the Chandra data (total clean exposure time 67.9ks) is shown in
Fig.~\ref{fig:xray}.  Overall, the pressure map shows that the cluster
is remarkably relaxed, although the localized region of high-pressure,
shocked gas in the southeast quadrant discovered and discussed by
\citet{komatsu01} and \citet{allen02} is also clearly visible (indicating recent merger activity).

The cluster mass profile was determined from the observed deprojected X-ray
temperature (Fig.~\ref{fig:xray}) and surface brightness profiles (see
\citealp{schmidt07} for details). The data from the southeast
quadrant (position angles of 180-280 degrees measured
west-through-north) were excluded from this analysis. In order to
compare with the lensing mass measurements presented in this paper, we
have projected the three-dimensional mass (and later also gas) profile
determined from the Chandra data, assuming that the cluster extends to
the virial radius (as determined from the X-ray data), $R_{\rm vir} =
3.03
\mbox{ Mpc}$. To estimate the systematic errors arising from this
choice of truncation radius, we adopt conservative limits and change
the radius by $30\%$. This resulted in changes to the projected masses
measured within 600kpc of the cluster center that are within the
quoted uncertainties; the precise value for the truncation radius is
therefore of limited importance for this study. The resulting
projected integrated mass profile is plotted in Fig.~\ref{fig:profile}
together with mass estimates obtained from lensing analysis.

\section{Cluster Mass Reconstruction from Strong and Weak Lensing Data}
 \label{sec:resultsw}
In this section we present the combined strong and weak lensing mass
reconstruction of {\rxj}. Here we essentially follow the method
described in \citet{bradac04a} and implemented on the ACS data in
\citet{bradac06}. The basic idea is to describe the cluster's
projected gravitational potential by a set of its values on a regular
grid $\psi_k$, from which we evaluate all quantities relevant for
gravitational lensing by finite differencing. E.g. the scaled surface
mass density $\kappa$ is related to $\psi$ via Poisson equation,
$2\kappa = \nabla^2\psi$ (where the physical surface mass density is
$\Sigma = \kappa \: \Sigma_{\rm crit}$ and $\Sigma_{\rm crit}$ is a
constant that depends upon the angular diameter distances between the
observer, the lens, and the source).

The advantage of such an approach is that we
avoid making any assumption on e.g. shape and/or profile of the
potential, which is crucial when dealing with merging clusters. The
strong and weak lensing data are then combined in a
$\chi^2$-fashion. We minimise the $\chi^2$ by searching for the
solution of the equation $\partial \chi^2 / \partial \psi_k =0$. We
linearize the resulting system (using the initial values for the
potential - see below - or values from the previous iteration) and
solve it using sparse matrix techniques. Since the weak lensing data
is noisy, regularisation (i.e. a process which ensures that unphysical
pixel-to-pixel variations in surface mass density are suppressed)
needs to be employed. For this purpose we compare the current surface
mass density map with one obtained on a coarser grid, and penalize
strong deviations in $\chi^2$ (see \citealp{bradac04a} for
details). The regularisation is chosen (and tested to be) such as to
not bias the resulting mass estimates.

\subsection{Initial conditions of the method}
For the purpose of obtaining the initial values of the potential
needed in the first step of the iteration we assume a single singular
isothermal sphere centered on the BCG with velocity dispersion $\sigma
= 1600\kms$, taken from the strong lens modeling. As noted by
\citet{eliasdottir07} there is a non-trivial conversion between $\sigma$
and $\sigma_{\rm PIEMD}$; therefore we simply assume a model with
similar enclosed mass within the Einstein radius.  We have chosen a
more simplistic model than in \S\ref{sec:slensing} on purpose, in
order not to bias our results and obtain a reconstruction independent
of the assumptions about the potential (as is the case in assuming
parametrised models). These particular assumptions, however, do not
influence the results much, when high quality data, such as the ones
here, are used (see discussion in
\S\ref{sec:syst}). Therefore any reasonable guess for a cluster potential (from X-rays, dynamical mass estimate, etc) can be used.

\subsection{Combined reconstruction}
As described in section~\ref{sec:slensing} we use 9 different multiply
imaged systems and ~700 weakly-lensed galaxies ($70\mbox{
arcmin}^{-2}$). We start with a $30 \times 30 \mbox{ pix}^2$ grid for
a $4.2\times 4.2
\mbox{ arcmin}^2$ field ($8\arcsec/\mbox{pix}$; the cluster is not centered in the ACS field) 
oriented with respect to the negative RA coordinate (not all of the
grid cells in the field contain data and those that do not are
excluded from the reconstruction). We gradually increase the number of
grid points in steps of 1, with the final reconstruction performed on
a $60 \times 60 \mbox{ pix}^2$  ($4\arcsec/\mbox{pix}$) grid.

\begin{figure*}[ht]
\begin{center}
\includegraphics[width=1.0\textwidth]{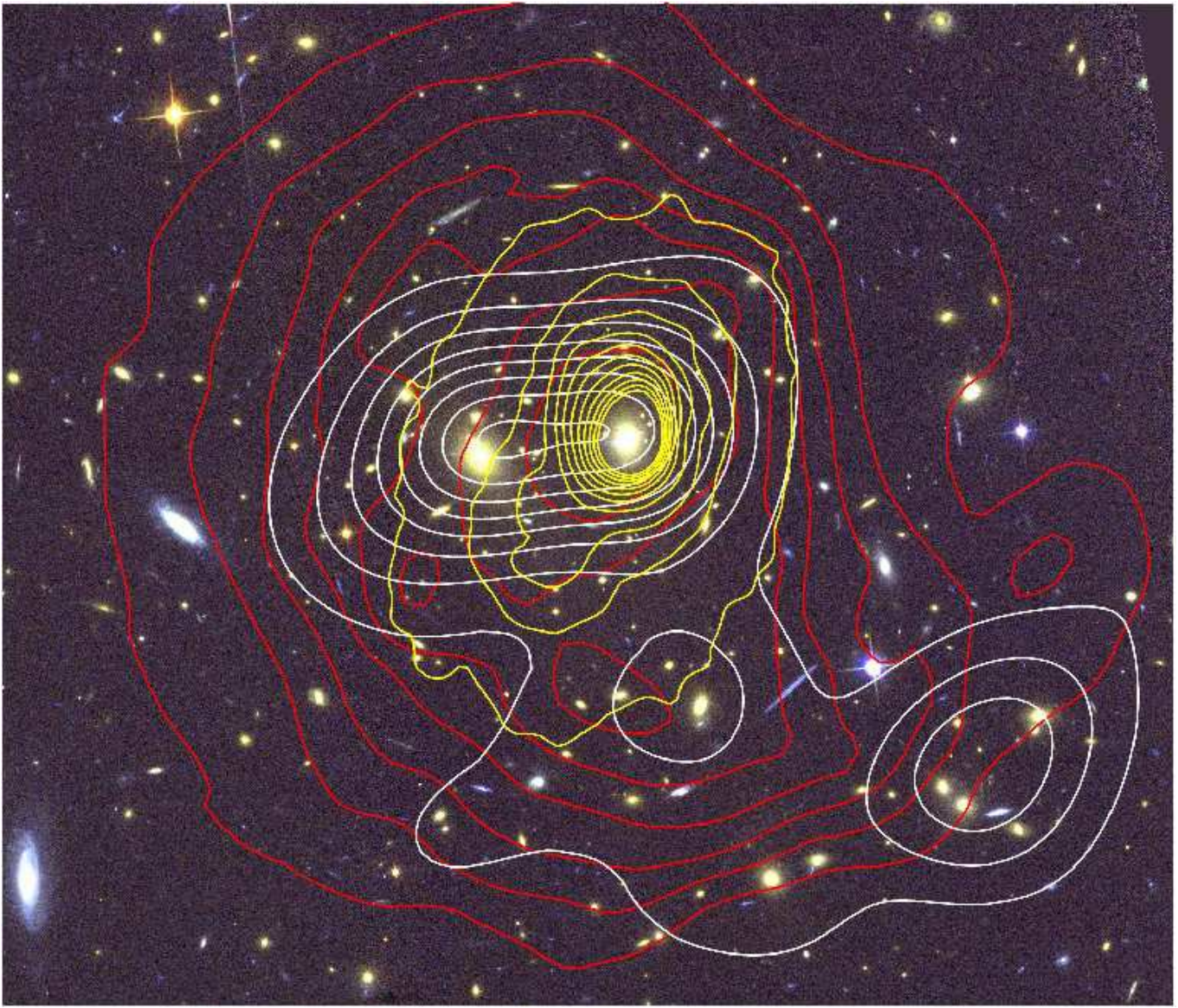}
\end{center}
\caption{The F450W-F814W-F850LP color composite of the cluster
  \protect \rxj. Overlaid in {\it red contours} is the surface mass
  density $\kappa$ from the combined weak and strong lensing mass
  reconstruction. The contour levels are linearly spaced with
  $\Delta\kappa = 0.2$, starting at $\kappa=0.5$, for a fiducial
  source at a redshift of $z_{\rm s} \to \infty$. The X-ray brightness
  contours (also linearly spaced) are overlaid in {\it yellow} and the
  K-band light (tracing the stellar mass) is overlaid in {\it
  white}. North is up and East is left, the field is
  $2.3^{\prime}\times 2.3^{\prime}$, which corresponds to $800 \times
  800 \mbox{ kpc}^2$ at the redshift of the cluster.}
\label{fig:swunited}
\end{figure*}

The resulting reconstruction is shown in Fig.~\ref{fig:swunited},
together with X-ray surface brightness contours from the Chandra
observation (see Sect.~\ref{sec:xraydata}). We clearly detect the main
cluster component, which is aligned with the BCG. The offset from the
BCG is $(-2\arcsec \pm 2\arcsec,-1\arcsec \pm 3\arcsec)$ and is also
consistent with the peak of the X-ray surface brightness. We also
resolve the south west structure, which also shows an overdensity of
cluster members. The resulting model predicts the strongly lensed
image positions with an average accuracy of less than $4\arcsec$,
which is also effectively the final pixel size. In order to reach
higher resolutions, adaptive grid methods needs to be employed; this will
be a subject of future work.

\begin{figure}[ht]
\begin{center}
\includegraphics[width=0.5\textwidth]{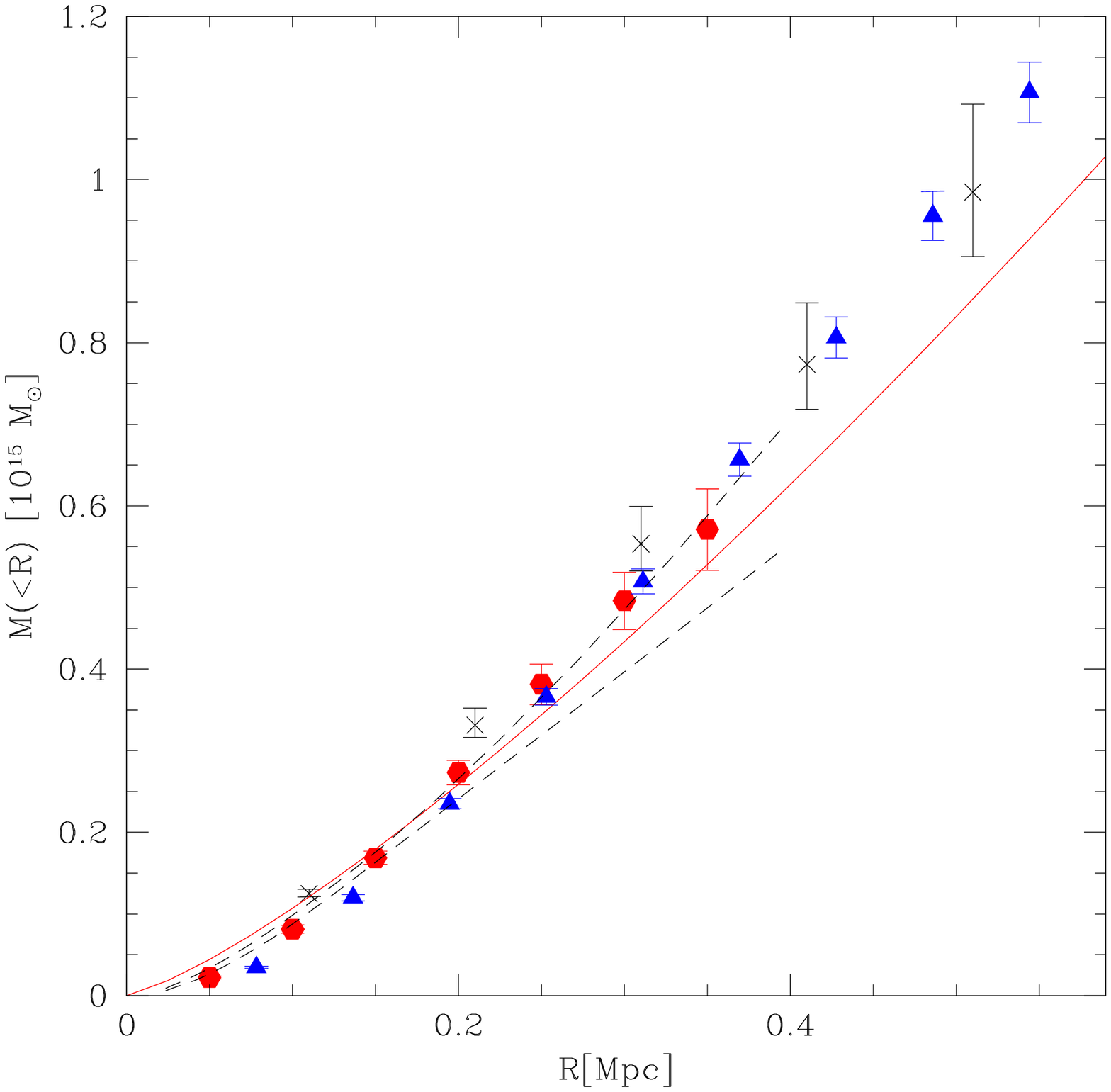}
\end{center}
\caption{The integrated, projected mass profile of \protect \rxj. The profile was
determined by measuring the enclosed mass in cylinders, centered on
the BCG  using strong and weak lensing (hexagons), strong lensing
using parametrised model (triangle) and X-ray data (crosses). We fit
power law profiles to the strong and weak lensing data, the result is
shown as a solid line. 1-sigma error region from the results obtained by \citet{halkola08} using only strong lensing data is shown as dashed lines.}
\label{fig:profile}
\end{figure}

The mass profile, calculated by determining the enclosed, projected
mass within circular apertures is plotted in
Fig.~\ref{fig:profile}. The errors include both systematic and
statistical contributions, their estimation is described in more
detail in Sect.~\ref{sec:syst}. We fit power-law model ($M(<R)
\propto R^{n}$) to the profile, and find it to poorly represent 
the data. The logarithmic slope of the best fit model is $n = 1.6 \pm 0.1$ (shallower than isothermal). Adopting an isothermal profile, we find a
line-of-sight velocity dispersions for $\sigma = (1550 \pm 100)
\kms$.

The total mass profile is in excellent agreement with the X-ray
data. It also agrees with the analysis in \citet{bradac04b}, and
strong lens modeling in \citet{halkola08} and the one presented in
\S\ref{sec:slensing} (the data points using only strong lensing
information in Fig~\ref{fig:profile} are extrapolated to a larger
radius using the adopted profile for modelling). The agreement between
our strong and strong and weak lensing modelling is expected, since
for the analysis in \S\ref{sec:slensing} we use the same set of
multiply imaged systems; \citet{halkola08} does use a subset of
multiple images, and the agreement between these semi-independent
analyses is very good. Most importantly, however, the agreement
between X-ray data and lensing analyses was not achieved to this
accuracy in the past and it is very encouraging to note that with
excellent data we can indeed measure cluster masses reliably.
Whereas the results from
\citet{bradac04b} show that even with the ground-based data one can
reliably measure enclosed mass, in order to obtain the full mass
profile space-based data are required. In addition, as shown by
\citet{halkola08}, if using a reasonable guess for the cluster
potential only a few systems are needed to reconstruct the mass
profile. However, further image systems, and the addition of weak lensing
data, increases the resolution and radial extent of the reconstruction,
which helps breaking the degeneracies in the mass profile as shown in
\S\ref{sec:dmbar}.

We also compare our estimates with the recent X-ray and gravitational
lensing comparison of \citet{gitti07}. The paper is based on old
lensing data (the ACS/HST data is not included) together with XMM
data. Whereas the lensing results have not changed significantly,
there is a disagreement between Chandra and XMM X-ray data.

The X-ray results presented here are in excellent agreement with those
reported by \citet{allen02} from a previous analysis of a Chandra data
set with shorter exposure. The X-ray derived masses are larger than
those reported from the analysis of XMM-Newton data by
\citet{gitti07}.  The origin of this discrepancy is likely to lie at
least in part in the complicating effects of the XMM-Newton point
spread function, which are significant in the presence of strong
temperature gradients (Fig.~\ref{fig:xray}) and which were not
modelled in detail by \citet{gitti07}.

The largest circular aperture we can place on the BCG is $350\mbox{
kpc}$. The enclosed, projected mass within this aperture is $M_{\rm
L}(<350\:\mbox{kpc})= (5.9 \pm 0.5) \times 10^{14} M_{\odot}$. The
uncertainties include statistical as well as systematic errors,
originating from weak lensing measurements, possible multiple image
and redshift misidentification, and the initial model we used. We now
describe the error budget in more detail.

\subsection{Errors and Possible Systematic Effects}
 \label{sec:syst} 

As in \citet{bradac06}, we also extensively study possible errors and
systematic effects. We generate 1000 bootstrap resampled weak lensing
catalogs and perform reconstructions on each of these.  To further
test the reliability of the strong lensing data we create 9 different
reconstructions, each time removing one of the multiply imaged systems
we use. The resulting $\kappa$-maps do not change substantially, the
main features (i.e. the ellipticity of the cluster and SW extension)
seen in Fig.~\ref{fig:swunited} remain in all of the
reconstructions.

We have also run the reconstruction by changing the average redshift
of weak lensing sources to $z_{\rm WL} = 1.0$ and $z_{\rm WL} =
2.0$. The changes in mass estimates were at the $1\%$ level; hence the
precise value of $z_{\rm WL}$ is unimportant in this case. The
normalisation of the mass profile is determined by the much stronger
signal coming from 9 multiple-image systems with ``assumed'' known
redshifts. Weak lensing is helping here with the shape of the mass
distribution in the areas where no multiply imaged systems are
present.

Finally we study the dependence of the results on initial
conditions. As noted in \citet{bradac06} the main features of the
reconstruction are independent of the initial conditions we
use. This is mostly attributed to the richness of the strong lensing
data used in that paper and here. Still, we performed mass
reconstructions where we changed the velocity dispersion of the initial
model to $\sigma = 1900\kms$ and $\sigma = 1300\kms$; the resulting
mass estimates are within the errors quoted above.

In summary, the errors on mass and surface mass density quoted
throughout this paper include the errors obtained from bootstrap
resampling the weak lensing catalogues, and removing individual strong
lens systems (the latter dominate the error budget). The other errors
discussed in this section have a minor contribution to the total error
budget.
\subsection{A Possible $z=4.08$ System?}
 \label{sec:z4} 
\begin{figure}[ht]
\begin{center}
\includegraphics[width=0.5\textwidth]{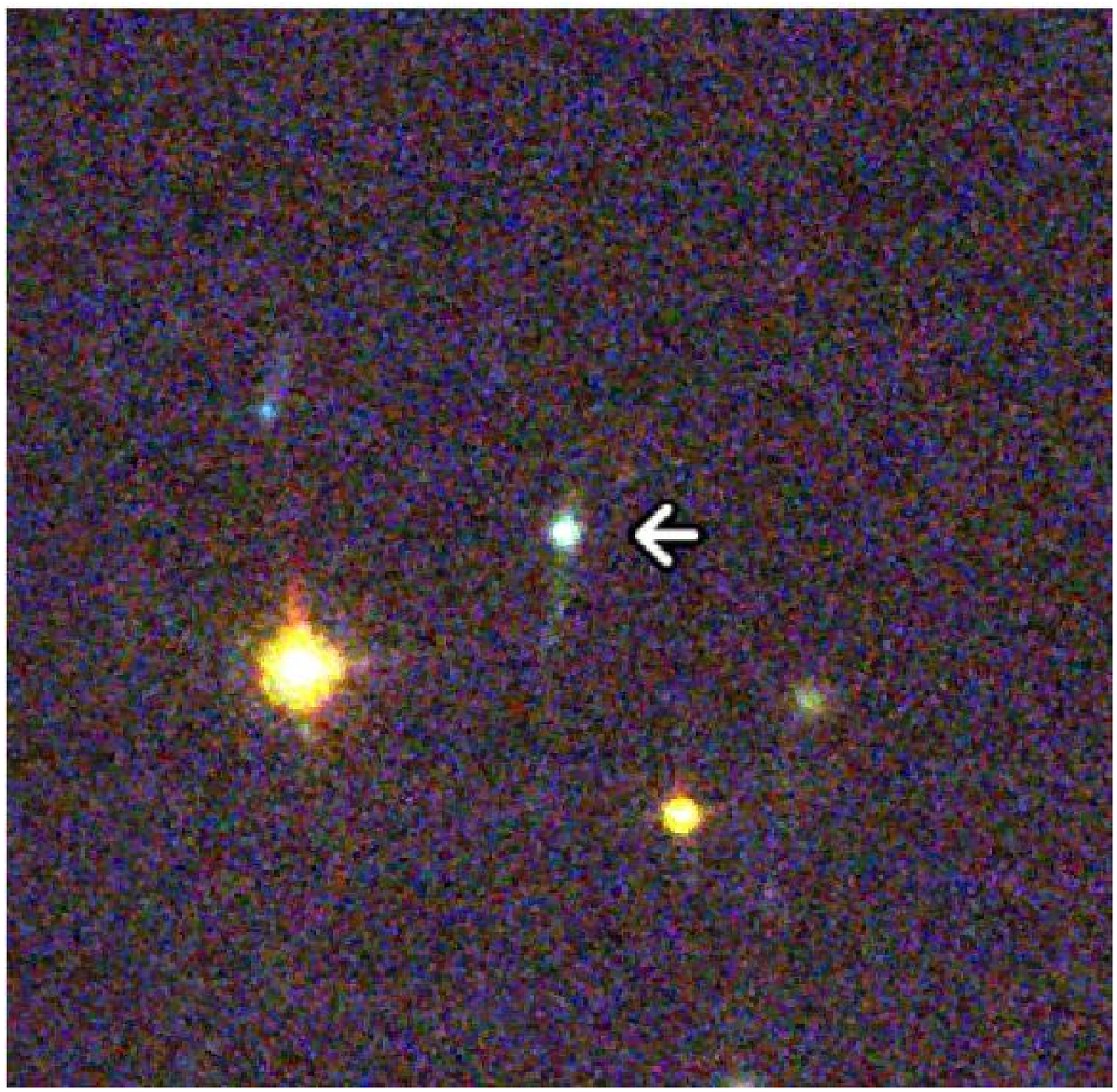}
\end{center}
\caption{$10\arcsec \times 10\arcsec$ cutout of the $z=4.08$ candidate discovered by \citet{cohen02} from the F450W-F814W-F850LP color composite.}
\label{fig:arcz4}
\end{figure} 
During their spectroscopy run, \citet{cohen02} serendipitously
discovered a candidate $z=4.08$ object (see
Fig.~\ref{fig:arcz4}). Using our final mass reconstruction we have
searched for counter images of this object, unfortunately without
success. If present we would have expected multiple images to be easy
to identify even in the absence of a reliable lens model, given its
somewhat distinctive colors (see Figure~\ref{fig:arcz4}).
Furthermore, the object is fairly bright in the F475W filter ($m_{\rm
F475W, AB} = 25.55 \pm 0.03$) and has a relatively blue F475W-F814W
color ($1.46 \pm 0.04$).  This new piece of information appears to be
inconsistent with the suggested redshift of $z=4.08$, since the flux
shortwards of Lyman alpha should be absorbed by the Lyman alpha forest
\citep{madau96}.

At present we can therefore neither confirm nor reject the possibility
of this object being lensed. However, we think it is unlikely that the
bright knot is indeed at a redshift of $4.08$. Since we also see some
underlying extended emission, it is possible that the bright knot is a
foreground object and the extended object is being lensed; due to its
low surface brightness we do not expect to be able to identify its
counter images. A more likely possibility is that the emission line
detected by \citet{cohen02} is the [\ion{O}{2}] doublet at 3727\AA,
implying a redshift of $z=0.66$ and therefore no multiple imaging. We
do not expect the doublet to be resolved given the spectral resolution
of $\sim10$\AA (FWHM). We conclude that the redshift of this object is
uncertain at present and further investigations are needed to either
confirm or reject the present redshift estimate.

\section{Dissecting RX~J1347.5$-$1145 into dark matter and baryons}
\label{sec:dmbar}

The relative distribution of dark matter and baryons in galaxy
clusters is important for a number of reasons. Whereas baryons are
only a minor total mass constituent in the dark matter dominated
clusters, in the centres of the clusters baryons are an important mass
component. In order to make a proper comparison with $\Lambda$CDM
predictions we need to disentangle both components. The
interplay between baryons and dark matter is important and can
potentially change the inner slope of dark matter halos \citep[see
e.g.][]{gnedin04}. Therefore in this section we separate the total
mass measurements we obtain from gravitational lensing analysis into
baryonic (gas and stars) and non-baryonic component and measure the
dark matter distribution of {\rxj}.

\subsection{2-D distribution of dark matter and baryons}
To estimate the stellar mass distribution we first estimate the
cluster K-band luminosity distribution by selecting the cluster
members using the photometric redshift (obtained using information
from 8 colors from ground-based data in \citealp{bradac04b}) cuts at
$[0.3-0.55]$. We measure their K-band luminosity by assuming the
absolute solar magnitude $M_{\rm K,\sun} = 3.28$, galactic extinction
following \citet{ned1} and
\citet{ned2} of $A_{\rm K}=0.023$, and K-correction from
\citet{poggianti97} of $K_{\rm K}(z_{\rm d}) = -0.25$.  We smooth the distribution using a Gaussian kernel of FWHM of
$30\mbox{kpc}$. To convert the stellar luminosity into stellar mass we
follow
\citet{drory04} and assume the stellar-mass-to-light ratio in K-band
to be $M_{*}/L_{\rm K}= 0.74\pm 0.3$ (see also
\citealp{bell03}).

The stellar K-band luminosity is shown in Fig.~\ref{fig:swunited} in
white contours. It follows the total mass distribution considerately
well, we detect the SW extension in both cases. Also shown is the
X-ray surface brightness, indicating that the major baryonic component
is spatially aligned with the distribution of the total mass and hence
dark matter within the uncertainties.

\subsection{Projected density profile of dark matter and baryons}
\begin{figure}[ht]
\begin{center}
\includegraphics[width=0.5\textwidth]{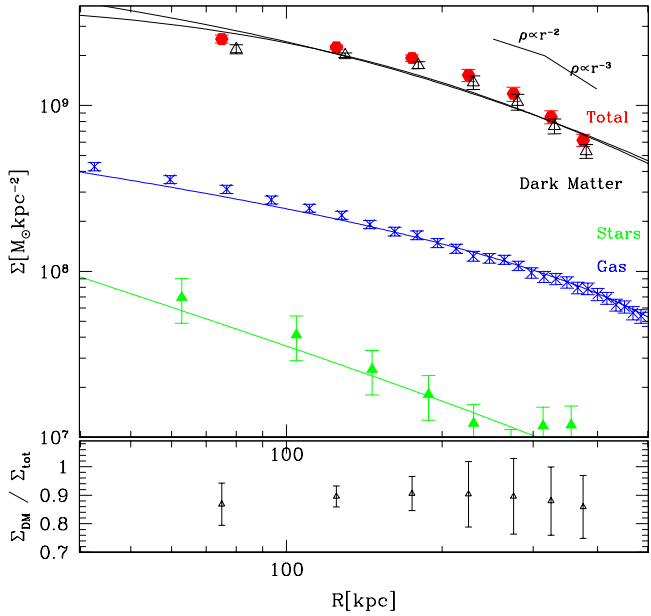}
\end{center}
\caption{(top) The projected mass density $\Sigma$ profile of {\protect \rxj} showing separately the stellar profile (filled triangles - green), the gas profile (stars - blue), dark matter profile (open triangles - black) and the total profile as measured from strong and weak lensing (hexagons - red). The dark matter profile has been fitted using the generalised NFW (cf. Eq.~\ref{eq:nfw}) - shown is the best fit model ($\beta = 0$, $r_{\rm s} = 160\mbox{kpc}$) and a best fit NFW model ($\beta = 1$, $r_{\rm s} = 350\mbox{kpc}$).  For gas and stars we use simple power law profiles. Also shown for reference are the $\Sigma$-profiles for 3D density profiles $\rho \propto r^{-2}$ and  $\rho \propto r^{-3}$. Dark matter points have been offset in the R-direction for clarity. (bottom) The ratio of dark matter to total matter.}
\label{fig:allprofile}
\end{figure}

In Fig.~\ref{fig:allprofile} we present the total surface (projected)
mass density profile $\Sigma$ from the lensing reconstruction. Further
we obtain the dark matter profile by subtracting the stellar and gas
density profile from the total density profile, as described next.

The stellar mass profile is obtained from the 2-D K-band light
distribution. We assume the dominant source of error comes from the
uncertainty of $M_{*}/L_{\rm K}$. We again smooth the distribution
using a Gaussian kernel of FWHM of $30\mbox{kpc}$; the exact value is
of lesser importance here, since stars contribute a minor fraction of
the total baryonic mass at radii $\gtrsim 50\mbox{ kpc}$ and we are
not sensitive in smaller scale variations in the total mass profile.

The projected gas profile is obtained from the observed X-ray surface
brightness profile and the deprojected X-ray gas temperature
profile. Finally, the dark matter profile is calculated from the
difference between the baryonic and the total mass profile from strong
and weak lensing mass reconstruction. In Fig.~\ref{fig:allprofile} we
show the individual contributions of stars, gas, and dark matter.

To estimate the inner slope of the dark matter halo we fit a
generalized, projected NFW profile: the 3D density is given by
\begin{equation}
\rho_{\rm DM}(r) = \frac{\rho_{0,\rm DM}}{(r/r_{\rm s})^{\beta}(1+r/r_{\rm s})^{3-\beta}}\; .
\label{eq:nfw}
\end{equation}
We determine the asymptotic inner slope $\beta$, the scale radius
$r_{\rm s}$ and the normalisation $\rho_{0,{\rm DM}}$ from the data
given in Fig.~\ref{fig:allprofile}.  We assume a flat prior $\beta >
0$, since negative values of $\beta$ give unphysical profiles. When
fitting the profile we include the full covariance matrix (i.e. taking
into account correlations between the bins) for the estimate of the
total density, gas density and the stellar mass density. The
covariance matrix was determined directly from strong and weak lensing
bootstrap resampled reconstructions and by sampling the models given
the uncertainties from gas and stellar mass measurements. We note
however, that another potential source of systematic errors will arise
from the lack of the symmetry in the mass distribution and somewhat
complicated geometry of this system. There is also the ambiguity when
choosing the center of mass for the profile estimation, which should
be computed around the center of mass. We chose to center the profile
on the BCG, which is a good approximation in our case, since the
maximum of $\Sigma$ is consistent with the position of the BCG
($2\arcsec \pm 4\arcsec$, i.e. $<10\mbox{kpc}$).

Furthermore, as already noted by \citet{sand07} there are strong
degeneracies between the generalised NFW parameters. In particular we
found a strong degeneracy between the scale radius $r_{\rm s}$ and the
inner slope $\beta$. The best fit profile gives $\beta=0.0 \pm 0.1$ and
$r_{\rm s} = 160 \mbox{ kpc} \pm 10\mbox{ kpc}$ ( $c_{200}=15$ and
$r_{200} = 2400 \mbox{ kpc}$). However, as shown in  Fig.~\ref{fig:allprofile} the NFW profile fit with a fixed inner slope $\beta=1$ is not significantly worse. The latter gives a scale radius of $r_{\rm s} = 350\mbox{ kpc}{\pm 100}\mbox{ kpc}$ ($c_{200}=6$ and $r_{200} = 2200 \mbox{ kpc}$).  The statistical
uncertainties alone on $c_{200}$ are of the order of $\sim 20\%$.

To reliably determine the concentration parameter and hence $\beta$ we
therefore need the data to extend to much larger scales (i.e. beyond
$r_{\rm vir}$). Strong lensing data alone (for this and also all the
other clusters) are not sufficient for this task. We plan to extend
this analysis by adding wide field weak lensing data that extend to a
radius of $\sim 30\arcmin$ ($10\mbox{ Mpc}$); this will be a subject
of a forthcoming paper.

Finally, the thermodynamic X-ray pressure map (Fig.~\ref{fig:xray})
shows the localized region of high-pressure, shocked gas in the
southeast quadrant. Therefore we are likely measuring a profile of two
post-merging, unequal mass clusters. This is further supported by the
presence of the eastern cD galaxy. Therefore the cluster might not be
fully relaxed and therefore the measurement of the total dark matter
distribution is unlikely to follow the NFW profile in great detail at
these small radii.

\section{Conclusions and Outlook}
 \label{sec:conclusions} The longstanding puzzle of the discrepant
mass estimates for the most X-ray luminous cluster seems to be
definitely resolved. Both our strong and weak lensing as well as X-ray
mass reconstructions agree well. In particular we draw the following
conclusions:
\begin{enumerate}
\item
Using the combined strong and weak lensing mass reconstruction we
derive a high-resolution, absolutely calibrated mass map; we get
projected, enclosed mass $M_{\rm L}(<350\:\mbox{kpc})= (5.9 \pm 0.5)
\times 10^{14} M_{\odot}$. Within the same radius the projected mass derived from X-ray data alone
is $M_{\rm X}(<350\:\mbox{kpc})= (6.6 ^{+0.6}_{-0.4}) \times 10^{14}
M_{\odot}$.
\item
The mass estimates are still in disagreement with previous dynamical mass
estimates by \citet{cohen02}. It is however possible that
\citet{cohen02} measured the velocity dispersion of the infalling
subcluster only \citep{allen02}, thereby biasing the mass estimate
low. If taken as a measurement of the cluster mass, such a
low velocity dispersion cluster is in complete disagreement with
strong lensing features we observe (regardless of their redshift). We
are currently analysing the remainder of spectroscopy data to resolve
this issue (Lombardi et al. in preparation).
\item 
Using the exquisite resolution of ACS data we are able to study the
spatial distribution of dark matter with respect to the baryons at
unprecedented accuracy. We clearly detect a mass concentration
centered on the BCG and a SW extension, which follows the light
distribution of the cluster members. Further we fit a generalised NFW
model to the dark matter density profile, finding strong degeneracies
between the inner slope $\beta$ and the scale radius $r_{\rm s}$. Our
data does not extend to sufficiently large radii, for more secure
determination of $\beta$, $r_{\rm s}$, and concentration parameter; we
plan to use weak lensing data extending beyond the virial radius in
the future to resolve this issue.
\end{enumerate}

The cluster {\rxj} is an extremely valuable object for understanding
the details of cluster formation and evolution. In addition, due to
their magnifying power, these most massive clusters are ideal tools to
study the high-redshift universe. By searching around the critical
line for high redshift sources as predicted by our lens model we were
already able to find high redshift multiple imaged systems. We plan to
extend this search in the future.

The large wavelength coverage of the data for this cluster has
proved to be extremely valuable for a detailed study of the cluster
mass distribution. With such data sets most massive
clusters of galaxies remain one of the key objects to constrain
cosmological parameters and to study the galaxy formation and
evolution from the early times until the present.


\begin{acknowledgements}
We would like to thank Masamune Oguri for many useful
discussions. Support for this work was provided by NASA through grant
number HST-GO-10492.03A from the Space Telescope Science Institute,
which is operated by AURA, Inc., under NASA contract NAS 5-26555.
This research has made use of data obtained from the Chandra Data
Archive and software provided by the Chandra X-ray Center (CXC). MB
acknowledges support from the NSF grant AST-0206286 and from NASA
through Hubble Fellowship grant \# HST-HF-01206.01 awarded by the
Space Telescope Science Institute. TT acknowledges support from the
NSF through CAREER award NSF-0642621, by the Sloan Foundation through
a Sloan Research Fellowship, and by the Packard Foundation through a
Packard Fellowship. This work was supported by the Deutche
Forschungsgemeinschaft under the project SCHN 342/3--1 and the German
Ministry for Science and Education (BMBF) under the projects
05AV5PDA/3 and 50 OR 0601. This project was also partially supported
by the Department of Energy contract DE-AC3-76SF00515 to SLAC.
\end{acknowledgements}

\end{document}